\documentclass[aps,prx,twocolumn,superscriptaddress]{revtex4-2}

\usepackage{amsmath,amssymb,amsthm}%
\usepackage{bm}%
\usepackage{stmaryrd}%
\usepackage{graphicx}%
\usepackage{hyperref}%
\usepackage{tcolorbox}
\usepackage[normalem]{ulem}

\renewcommand{\vec}[1]{\boldsymbol{ #1 }}

\newcommand{\mm}[1]{{{#1}}}
\newcommand{\mmnew}[1]{{{#1}}}
\newcommand{\jk}[1]{{{#1}}}
\newcommand{\sgd}[1]{{{#1}}}
\newcommand{\sgdnew}[1]{{{#1}}}

\newcommand{\st}[1]{{{#1}}}
\newcommand{\rp}[1]{{{#1}}}
\begin{document}

\title{A Driven Disordered Systems Approach to Biological Evolution in Changing Environments}

% Use letters for affiliations, numbers to show equal authorship (if applicable) and to indicate the corresponding author
\author{Suman G Das}
\affiliation{Institute for Biological Physics, University of Cologne, Z{\"u}lpicher Stra{\ss}e 77,
D-50937 K{\"o}ln, Germany}
\author{Joachim Krug } 
\affiliation{Institute for Biological Physics, University of Cologne, Z{\"u}lpicher Stra{\ss}e 77,
D-50937 K{\"o}ln, Germany}
\author{Muhittin Mungan}
\affiliation{Institute for Biological Physics, University of Cologne, Z{\"u}lpicher Stra{\ss}e 77,
D-50937 K{\"o}ln, Germany}
\affiliation{Institut f{\"u}r Angewandte Mathematik, Universit{\"a}t Bonn, Endenicher Allee 60, D-53115 Bonn, Germany}


\begin{abstract}
  Biological evolution of a population is governed by the fitness landscape, which is a map from genotype to fitness. However, a fitness landscape depends on the organism’s environment, and evolution in changing environments is still poorly understood. We study a particular model of antibiotic resistance evolution in bacteria where the antibiotic concentration is an environmental parameter and the fitness landscapes incorporate tradeoffs between adaptation to low and high antibiotic concentration. With evolutionary dynamics that follow fitness gradients, the evolution of the system under slowly changing antibiotic concentration resembles the athermal dynamics of disordered physical systems under external drives. \st{Exploiting this resemblance, we show that} our model can be described as a system with interacting hysteretic elements. \st{As in the case of the driven disordered systems, adaptive evolution under antibiotic concentration cycling is found to exhibit} hysteresis loops and memory formation. We derive a number of analytical results \sgdnew{for quasistatic concentration changes.} \mmnew{We also perform numerical simulations} \sgdnew{to study how these effects are modified under driving protocols in which the concentration is changed in discrete steps.}  
%  \st{We derive a number of analytical and numerical results.}
  Our approach provides a  general
  framework for studying motifs of evolutionary dynamics in biological systems in a changing environment. 
  %% MM 171 words, PRX guidelines are abstracts less than 500 words in length. 
\end{abstract}

%\dates{This manuscript was compiled on \today}
%\doi{\url{www.pnas.org/cgi/doi/10.1073/pnas.XXXXXXXXXX}}

\maketitle
%\thispagestyle{firststyle}
%\ifthenelse{\boolean{shortarticle}}{\ifthenelse{\boolean{singlecolumn}}%{\abscontentformatted}{\abscontent}}{}
\section{Introduction}
The concept of the fitness landscape \st{of a biological species}, introduced by Sewall Wright \cite{wright1932roles}, is a useful tool for \st{understanding} \rp{evolutionary processes. According to this picture, evolving populations are driven uphill along fitness gradients by natural selection.}
Mathematically, a fitness landscape is a map \rp{which assigns fitness values to genetic sequences.}
% \mm{which assigns a fitness value to each genotype of a species}.
In recent decades, it has become 
\st{feasible} to empirically determine fitness landscapes \st{comprising several mutations}, and a wealth of new work
has illuminated various aspects of evolutionary dynamics on \rp{different classes of} fitness landscapes \cite{weinreich2006darwinian,
  %depristo2007mutational,marcusson2009interplay,
  lozovsky2009stepwise,
  %brown2010compensatory,
  schenk2013patterns,de2014empirical,palmer2015delayed,bank2016on,domingo2018pairwise,fragata2019evolution,pokusaeva2019an}.
%goulart2013designing,knopp2018predictable}.
\st{At the same time, the availability of empirical data has renewed the interest in studying fitness landscapes theoretically} 
% {\color{red}The theoretical study of fitness landscapes
% and its application to empirical data
% have also experienced increased activity in recent decades} 
\citep{franke2011evolutionary,szendro2013quantitative,
  % weinreich2013should,
  neidhart2014adaptation,ferretti2016measuring,blanquart2016epistasis,crona2017inferring}.
% ,crona2020rank
% and the application of ideas from statistical mechanics to these problems has been particularly useful
% \cite{sella2005application,koonin2011logic,manhart2014statistical,hwang2018universality}.

A less well-studied topic in this field is evolution in \st{{\it changing}} environments. Fitness landscapes are a function of environment, and can change in systematic ways as environmental parameters change. Whereas the fitness landscape provides information about $G \times G$ (gene-gene) interactions, the introduction of the environmental parameter furnishes information about $G \times G \times E$ (where $E$ stands for environment) interactions, i.e., about how the environment modifies the  gene-gene interactions
\cite{murugan2021roadmap,deVos2018ecology,gorter2018local,anderson2021adaptive}. A few studies on microbial growth have measured or interpolated fitness values as a function of environmental parameters \cite{mira2015adaptive,ogbunugafor2016adaptive,das2020predictable}, but systematic theoretical work in this field is still limited.

\begin{figure}
\centering
 \includegraphics[width=6cm,height=4cm]{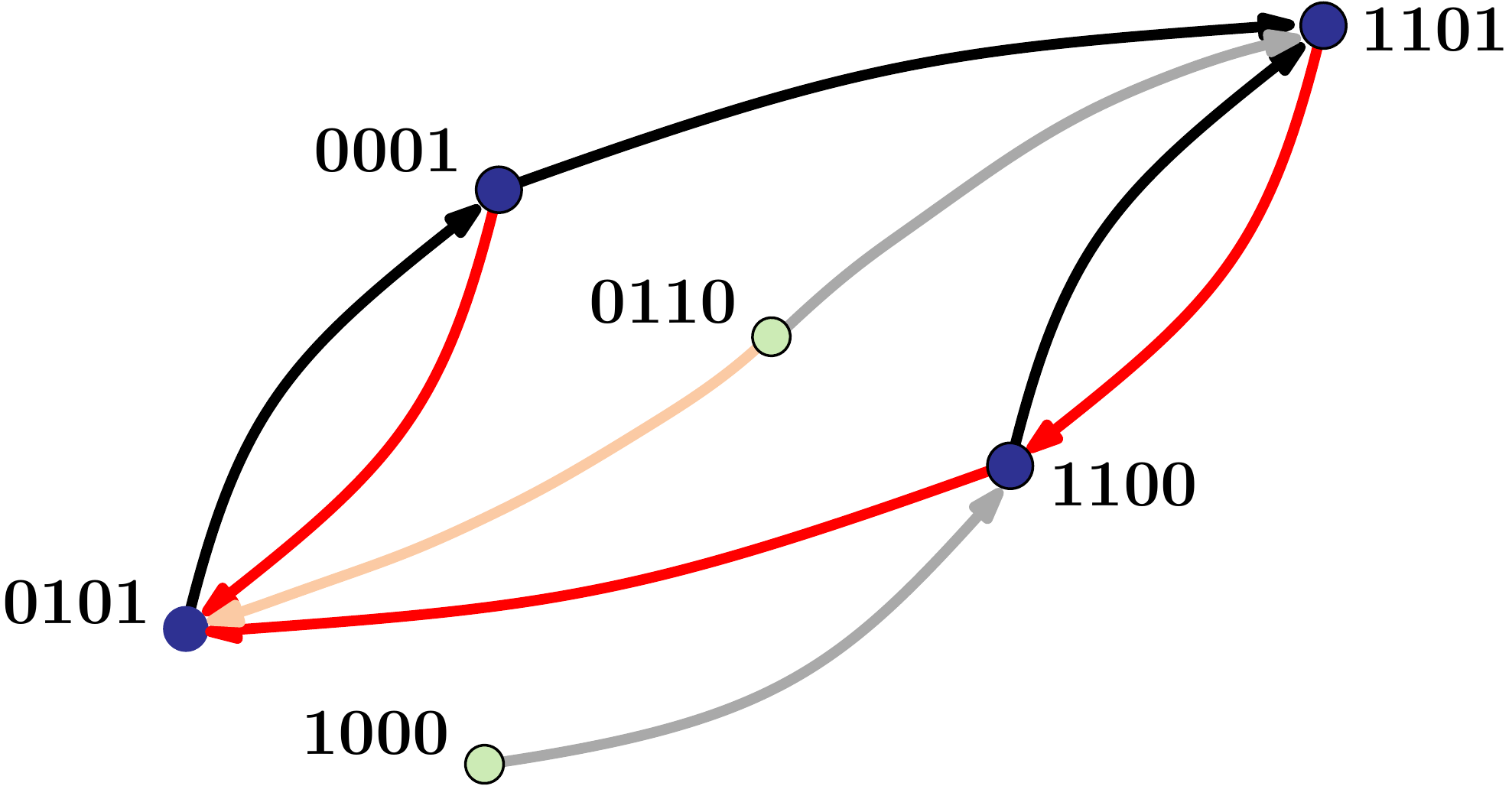}
 \caption{\textbf{State transition graph of antibiotic resistance evolution.} The nodes depict genotypes composed of four mutations
   in the antibiotic resistance enzyme TEM-50 $\beta$-lactamase. Genotypes are represented as binary strings where a $1$ denotes the presence and $0$ the absence of a specific mutation. The growth rates of bacteria expressing these mutant enzymes were reported in \cite{mira2015adaptive} for the antibiotic piperacillin
   % (with a fixed concentration of a $\beta$-lactamase inhibitor)
   at three different concentrations ($128$ $\mu$g/ml, $256$ $\mu$g/ml and $512$ $\mu$g/ml). Each node is a local fitness maximum
   at one of these concentrations.  
   Black and grey arrows connect nodes that would be reached under adaptive evolution when the concentration is increased, and red and orange arrows represent the dynamics under concentration decrease. For example, $0001$ is a local maximum at $256$ $\mu$g ml, but when the concentration is switched to $512$ $\mu$g/ml, it is no longer a fitness maximum. Evolution through a greedy adaptive walk \sgd{(where every step is maximally fitness increasing)}
   leads to the new maximum $1101$. The graph displays a hysteresis loop $0101 \to 0001 \to 1101 \to 1100 \to 0101$. \mm{The green nodes are transient and cannot be reached under cyclic concentration changes. \mm{Gray and orange arrows mark} transitions out of transient states.}  
 }
 \label{fig:mira}
\end{figure}

Understanding  
and predicting the effect of the environment on fitness landscapes has important practical applications.
A pertinent example is the case of antibiotic resistance in bacteria, where it has been shown that the fitness landscape depends strongly on the antibiotic concentration \cite{mira2015adaptive,ogbunugafor2016adaptive}. Uncontrolled variation in antibiotic concentration, both in clinical settings and elsewhere  \cite{kolpin2004urban,andersson2014microbiological}, is a cause for the rise in antibiotic resistance, which is a major clinical challenge today. Figure \ref{fig:mira} shows an empirical example
% from \cite{mira2015adaptive}
of the kind of processes we are interested in. The \rp{fitness} values of the genotypes \rp{(i.e. genetic sequences)} in the figure were measured in \cite{mira2015adaptive}, and based on \rp{them}, one can predict transitions between genotypes under concentration increase (black/gray arrows) or decrease (red/orange arrows). Notice that this small system already exhibits some interesting properties, such as a hysteresis loop under antibiotic concentration cycling and transient genotypes that are not part of the loop.

Our focus is primarily on \st{antibiotic resistance evolution,}
% one such class of problems, 
where the environmental dependence of the fitness landscape is governed by a
tradeoff between two phenotypes, bacterial growth rate and resistance \cite{das2020predictable}. While we will mostly use the language of antibiotic resistance evolution in the following,
% two sections,
the theory developed is more generally applicable, as will become clear from  the mathematical model.
% in the next section.
Our work uses tools from statistical physics, specifically the physics of disordered systems. Concepts and methods from
statistical physics have been used in the theory of evolution for a long time \cite{fisher1958genetical,koonin2011logic}.
% ; see Chapter 4 of \cite{koonin2011logic} for a broad overview of the topic.
Precise quantitative analogies with evolutionary phenomena have been found with equilibrium statistical physics \cite{sella2005application}, the theory of random walks \cite{manhart2014statistical}, spin glasses
% \cite{schmiegelt2014evolutionary}
\cite{hwang2018universality,stein1992spin,franz1993evolutionary}, and many more.
Most of these, however, focus on static fitness landscapes.

Here we investigate evolution on rugged landscapes, i.e. landscapes with a large number of local fitness maxima, that {vary} with changes in an external parameter. This setting is naturally reminiscent of the physics of driven disordered systems, 
particularly in the athermal quasistatic (AQS) regime \cite{maloney2006amorphous}, where thermal activation processes 
are absent or negligible. The primary effect of the external forcing is then to alter the set of stable equilibria or their locations. As a result, under a time-varying external forcing such systems remain in a given equilibrium until it becomes unstable, and a fast relaxation process leads to a new equilibrium. Despite the absence of thermal activation processes, the resulting dynamics can
nevertheless be rather complex, exhibiting memory effects \cite{keim2019memory,mungan2019networks} as well as dynamic phase transitions, such as the jamming transition in granular materials \cite{behringer2018physics}, or the yielding transition in amorphous solids \cite{bonn2017yield}.

\st{In particular, we} find that evolutionary genotypic change has close parallels with systems such as cyclically sheared amorphous solids \cite{regev2013onset,fiocco2013oscillatory}, where a changing environmental parameter is analogous to an external shear, and transitions to new genotypes are similar to {localized} plastic events inside the solid \st{which can exhibit hysteresis}. As was shown recently \cite{munganterzi2018,munganwitten2018,mungan2019networks}, the AQS conditions permit a rigorous description of the dynamics of such systems in terms of a directed
\textit{state transition graph}.
% the {\em AQS transition graph}.  
Since the transition graph represents the response of the system to any possible deformation protocol, it provides a bird's-eye view of the possible dynamics % Thus the dynamics is encoded in the topology of the transition graph and the latter can be used to infer information about the former, as was done recently in the context of the sheared amorphous solids
\cite{mungan2019networks,regev2021topology,keim2021multiperiodic}.

\rp{The main goal of this paper is to show that the driven disordered systems approach leads to new insights into evolution in changing environments, such as the prevalence of hysteresis, precise rules for the substitution of mutations along fitness-increasing paths in the rare-mutation regime, and the encoding of the evolutionary past in the genome. We establish new quantitative results, such as the number of fitness maxima across the entire permissible range of environmental parameters, the mean number of mutations that fix after an instability occurs, and the extent of reversibility of the adaptive evolution.} 
\mmnew{Finally, we extend our results by going beyond the quasistatic limit, considering the adaptive response to discrete jumps in concentration. We find that} \sgdnew{several of our results carry over to this more realistic setting, but there are interesting differences as well when large jumps in concentration are involved.}  

 \rp{Our analysis is carried out on a model of antibiotic resistance evolution. This model is based on empirical observations on the generic properties of dose-response curves obtained in the literature on drug resistance and
provides a principled way of describing the environment-dependence of fitness landscapes. This introduces a new dimension into the traditional study of fitness landscapes, which has mostly been concerned with a fixed environment.
Exploiting the analogy with disordered systems, we find that the transition graphs describing the evolution of antibiotic resistance have a structure that bears strong resemblance to the Preisach model \cite{preisach1935magnetische} of hysteresis in magnets, but in a generalized setting where the 
elementary units of hysteresis interact with each other \cite{hovorka2005onset,keim2021multiperiodic,lindeman2021multiple,hecke2021profusion,bense2021complex}.
Lastly, we believe that the disordered systems view point, in particular the state transition graph approach, is a useful addition to the mathematical repertoire of evolutionary theory.}        

\sgd{The paper is structured as follows. In Section II, we introduce
  the mathematical model and the central concepts essential to its
  analysis. Section III contains the results and is divided into seven
  subsections\mm{. S}ubsections A-C develop the formalism of state
  transition graphs\mm{ and construct the set of local fitness maxima. Here we also} derive several general results related to
  hysteresis and memory formation under quasistatic environmental
  change using this formalism\mm{. S}ubsection D provides statistical
  results regarding the number of local fitness maxima across
  environments\mm{, while} subsection E reports numerical results on phenotypic
  reversibility, with statistical averages performed over evolutionary
  trajectories
  % with natural random walks
  with quasistatically changing concentration\mm{. S}ubsection F goes beyond the quasistatic approximation by studying dynamics under discrete changes in concentration. Subsection G focuses on reversibility at the genotypic level and compares and contrasts it with phenotypic reversibility. Section IV summarizes the main results and discusses future directions for research.} 

\section{Model}
We define a genotype $\vec{\sigma}$ as a binary string of length $L$, i.e. $\sigma_i \in \{0,1\}$, where $i=1,2,\dots,L$ denotes the sites where mutations can occur, and $\sigma_i=1$ indicates the presence of a mutation. An equivalent and useful way of thinking about $\vec{\sigma}$ is as a set of mutations drawn from a total of $L$ {possible}
%different 
mutations. The genotype without mutations, commonly
referred to as the \textit{wild type}, is then the empty set, whereas the \textit{all-mutant} is the set with all the $L$ mutations.  We will use the notation $\vec{\sigma}$ both as a string and as a set, and clarifications on the notation will be provided wherever necessary.
%A mutation is the analogue of a hysteron in a Presisach model.

\begin{figure}%[\sidecaptionrelwidth][t]
\centering
\includegraphics[width=8.6cm,height=10.5cm]{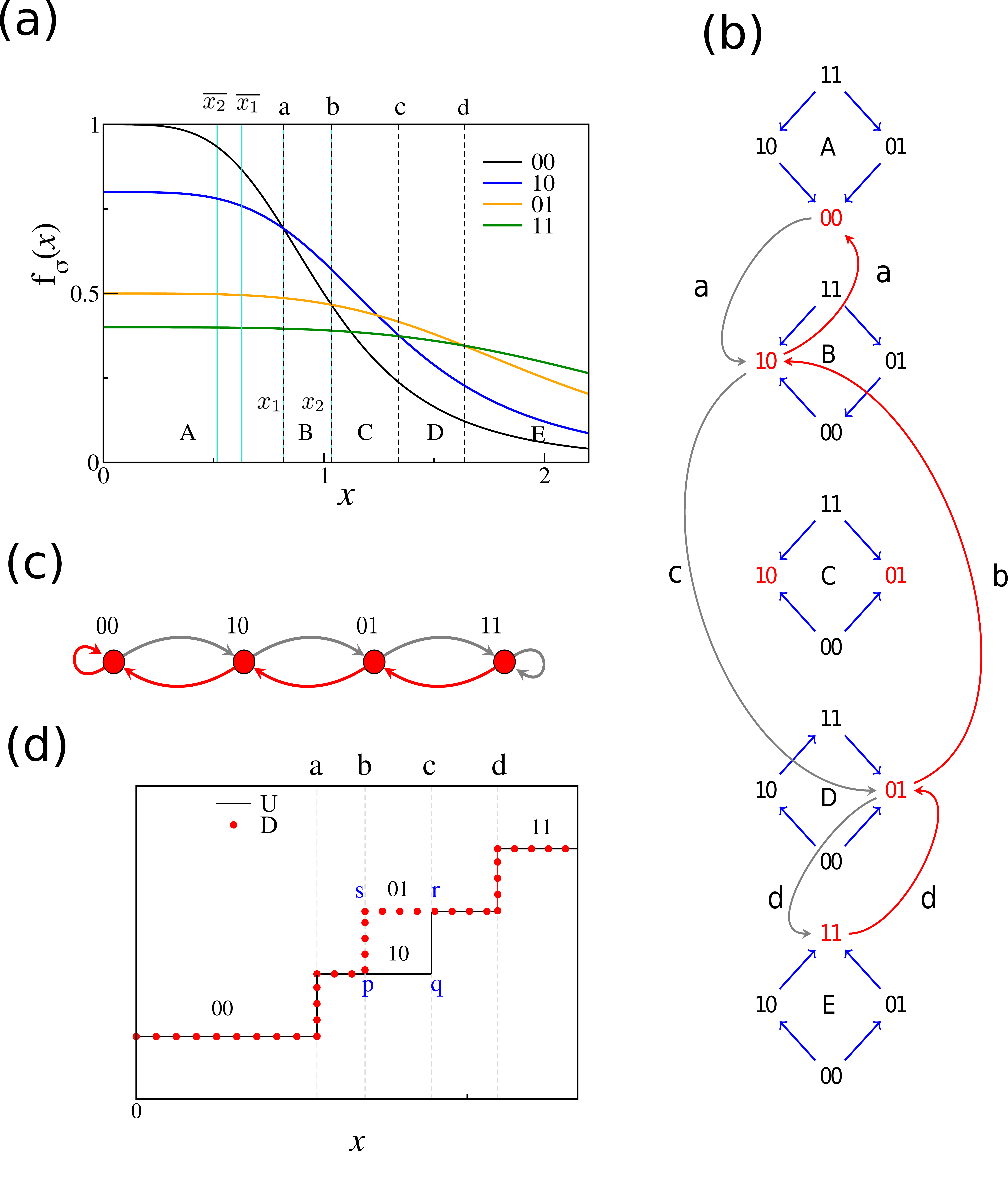}
\caption{\textbf{Tradeoff-induced fitness landscapes (TIL)
    model.} ({\bf a}) Fitness curves for four genotypes in a TIL model
  with two sites ($L=2$), with parameters $r_1=0.8,~ m_1=1.3$ and 
  $r_2=0.5,~m_2=2$. The shape of the curve is the Hill function $w(x)=1/(1+x^4)$.
  % i.e. the growth rate of bacterial cultures as a function of drug concentration $x$
  \cite{chevereau2015quantifying,regoes2004pharmacodynamic}.
The figure is divided into five regions A-E, corresponding to different fitness graphs. A new fitness graph occurs when the fitness curves of two mutational neighbors intersect. The $x$-values of the intersection points are marked by the letters a-d. { The \mm{elements $x_1$}, $x_2$, $\overline{x_1}$, $\overline{x_2}$  of the ordering sequence  (see main text) are indicated by solid vertical lines.} ({\bf b}) Fitness graphs in the regions A-E. Concentration \st{$x$} increases in the downward direction. In each fitness graph, the local fitness maxima (LFMs) are marked in red. Evolution in a fitness graph follows the oriented edges until a fitness maximum is reached. The curved grey arrows follow the evolution of the system under quasistatic increase of $x$ starting from the stable state $00$ at $x=0$ until the all-mutant $11$ is reached; the curved red arrows continue the trajectory as the concentration is quasistatically decreased until $00$ is reached again. 
%The points on the $x$-axis at which the transitions occur are stated next to the arrows. 
%Note that not all curved arrows involve changes at a single site. The transition at c alters both sites in two steps, $10 \to 11 \to 01$. 
({\bf c}) Transition graph for the two-site system is shown. { This should be distinguished from the fitness graphs in panel (b).} The nodes of the transition graph are the stable states which, in this simple case, comprise all genotypes. The {grey} arrows are the U transitions, i.e. the transitions under concentration increase, and the {red} arrows are the D transitions, i.e. transitions under concentration decrease. The transition graph can be read off from the sequence of fitness graphs in panel (\textbf{b}). ({\bf d}) The transition \st{between} states is shown schematically. Each horizontal level is a genotype, and the vertical lines denote transitions. The black lines correspond to genotypes reached under U transitions (starting from $00$ at $x=0$), while the line traced out by the red dots indicates the genotypes reached under D transitions (starting from $11$ at large $x$).
The hysteresis loop {\it pqrs} is marked in the figure.}
\label{fig:intro}
\end{figure}

Our focus is on the tradeoff induced landscapes (TIL) model introduced in \cite{das2020predictable}, which is defined through three key properties that
are motivated by empirical observations. 1) The fitness of each genotype {$\vec{\sigma}$} is a function of an environmental parameter $x\ge 0$, 
\st{the antibiotic drug concentration, and is described by a fitness curve of the form} 
\begin{equation}
f_{\vec{\sigma}} = r_{\vec{\sigma}} w(x/m_{\vec{\sigma}}).
\label{eq:scaling}
\end{equation}
The fitness curve thus has the same shape for different genotypes
except for a rescaling of the axes by the genotype-specific parameters
$r_{\vec{\sigma}}$ and $m_{\vec{\sigma}}$. {This is a common observation for various bacterial strains and antibiotics \cite{chevereau2015quantifying,das2020predictable,lukavcivsinova2020highly}}.
{We call $r_{\vec{\sigma}}$ the null-fitness and $m_{\vec{\sigma}}$ the resistance of a genotype $\vec{\sigma}$, following terminology used for bacterial \textit{dose-response curves} that represent the population growth rate as a
  function of drug concentration \cite{das2020predictable,regoes2004pharmacodynamic,chevereau2015quantifying}.
  % We refer to $r_{\vec{\sigma}}$ and $m_{\vec{\sigma}}$ as marginal phenotypes of the genotype $\vec{\sigma}$.
  We choose units such that for the wild type $\vec{\sigma} = \vec{0}$, $r_{\vec{0}} =1$ and $m_{\vec{0}}=1$, so that 
%MM: I think it is easiest if we refer to genotypes using the sigma notations, i.e. \vec{0} for WT and \vec{1} for the all mutant
% where the subscript $wt$ denotes the wild type. 
  $f_{\vec{0}}(x)=w(x)$}.  Further, $w(x)$ is a monotonic decreasing function, reflecting the decreasing fitness of a bacterial cell with increasing drug concentration. { For numerical purposes, we will choose the widely-used Hill function form: $w(x)=1/(1+x^n)$, where $n$ is called the Hill exponent. Empirically obtained dose-response curves are frequently fitted through Hill functions \cite{regoes2004pharmacodynamic}, and the scaling property expressed in \eqref{eq:scaling} shows up as a common value of $n$ shared by various mutants of the same strain exposed to the same drug \cite{chevereau2015quantifying,lukavcivsinova2020highly}.} 2) Every mutation comes with two parameters $r_i$ and $m_i$, and for any genotype, {$r_{\vec{\sigma}} = \exp[\sum_i \sigma_i \ln r_i]$} and $m_{\vec{\sigma}} = \exp[\sum_i \sigma_i \ln m_i]$. Thus, the effects of individual mutations combine
% without epistasis, i.e
in a simple multiplicative manner. { This is based on empirical observation that phenotypes of null-fitness and resistance in fact exhibit limited or no epistasis for several microbial species and drugs \cite{marcusson2009interplay,das2020predictable,knopp2018predictable}.} 3) { The mutations exhibit tradeoff between adaptation to low and high drug concentrations \cite{andersson2014microbiological,marcusson2009interplay,melnyk2015fitness}}, { i.e.} $r_i < 1$ and 
$m_i>1$. This means that every mutation enhances the resistance, but this comes at the cost of reduced null-fitness.
The fitness curves of a specific {realization of the} TIL model with $L=2$ mutations are shown in Fig \ref{fig:intro}(a).

The problem of {analyzing} this model has two components. First, one needs to understand the topography of the fitness landscape, \st{i.e.} the set of local fitness maxima and the paths that lead to the maxima, for a fixed $x$. The second part involves questions about evolutionary dynamics between maxima under 
changing drug concentrations. The first part has been addressed in detail in
\cite{das2020predictable}, and we describe some of the salient
features of landscape topography here. The landscape of the TIL model
is highly rugged (except at very low and very high $x$), i.e. the
number of fitness maxima is asymptotically exponential in $L$
\cite{das2020predictable}. To describe evolutionary dynamics at fixed
$x$, it is useful to introduce the notion of a {\it fitness graph}. The nodes of the { fitness } graph are the genotypes, and edges connect \st{{\it mutational neighbors}}, i.e. genotypes that differ by a single mutation. The
fitness graph is an acyclic \jk{oriented} graph, where the edges point towards
increasing fitness \cite{crona2017inferring,crona2013the}. 
\sgd{The fitness graph depends on $x$: a
fitness maximum for a certain value of $x$ may not be a fitness
maximum for another (see Fig. \ref{fig:intro}(b) for an example).
Note that the fitness graphs change only when the fitness curves of two mutational neighbors intersect.}
Evolution is assumed to
proceed through adaptive walks, i.e. the entire population moves along
the edges of the fitness graph respecting their orientation
\cite{hwang2018universality,kauffman1987towards,orr2002population,seetharaman2014adaptive,agarwala2019adaptive} \sgd{(see Appendix A for further details)}.  \sgd{While this is an idealization, adaptive walks have been found useful in the analysis of microbial evolution experiments \cite{rokyta2005empirical, rokyta2009genetics, schoustra2009properties}.}

Along the path taken by an adaptive walk, the fitness
increases monotonically, and such paths are \st{therefore} {\it(evolutionarily) accessible}
\cite{weinreich2006darwinian,franke2011evolutionary,weinreich2005perspective}. The
adaptive walk terminates once a local fitness maximum is reached. In general, there are multiple accessible paths starting from a genotype. 
A {\em greedy adaptive walk} is an adaptive walk {in} which every step is maximally fitness increasing
\cite{kauffman1987towards}. \jk{A more realistic dynamics is obtained by assuming that the probability of a transition $\vec{\sigma} \to \vec{\sigma}^\prime$ is proportional to $1-e^{-2s}$ when $s>0$ and $0$ otherwise, where $s\equiv f_{\vec{\sigma}^\prime}/f_{\vec{\sigma}}-1$
  denotes the selection coefficient \cite{orr2002population,seetharaman2014adaptive}.
% \sgd{A more realistic scenario is described by a {\em natural adaptive walk}, where the probability of a transition $\vec{\sigma} \to \vec{\sigma}^\prime$ is proportional to $1-e^{-2s}$ when $s>0$ and $0$ otherwise \cite{kimura1962probability}; the selection coefficient $s\equiv f_{\vec{\sigma}^\prime}/f_{\vec{\sigma}}-1$.
We have used this version of the adaptive walk dynamics for all simulations, and the greedy walk has been used in the discussion of some topological properties of specific transition graphs.}

% of the fitness graphs generated by a TIL model).
We also make use of the
notion of {\it mutationally directed} (or simply {\it directed})
paths, which are paths in the fitness graph along which the number of
mutations \mm{{\em relative to the wild type}} increases or decreases monotonically. The following interesting property about the TIL landscapes at fixed $x$ was
established in \cite{das2020predictable}. It is worth discussing, since we will use it to prove certain results in the following sections. \\
\noindent{\it Directed Path Accessibility: \mm{Given a fixed concentration $x$}, every (mutationally) directed path 
ending at a local  fitness maximum is accessible.}\\ 
In other words, every local maximum $\vec{\sigma}$ is evolutionarily accessible from the subsets (respectively supersets) of $\vec{\sigma}$ by a sequential gain (resp. loss) of mutations; the mutations may be gained (or lost) in any order. This property has remarkable consequences. 
For example, the wild type is a subset of every genotype, and therefore can access every fitness maximum through all
{{directed}} paths. Whenever the  wild type is a fitness maximum, it must be the only fitness maximum in the landscape, since it can be accessed from all genotypes. The same two properties also hold for the all-mutant.
With this background, we move on to the main focus of this article, which is evolutionary dynamics under slow changes in $x$. It is here that the 
%analogy with
{relation with} \st{driven disordered systems, in particular the Preisach model,} will become apparent.

\section{Results} 
\subsection{Stable states}
We consider an evolutionary dynamic \sgd{where the system is driven by changing the parameter $x$, and at each value of $x$ we wait long enough so that the system reaches a local \mm{fitness} maximum}
% at $x$
through an adaptive walk. \sgd{An estimate of the necessary waiting time in terms of population-genetic parameters is provided in Appendix A.} We call a genotype a {\em stable state} if it is a local fitness maximum \mm{(LFM)}
at {\it some} concentration $x$.
 As $x$ changes, the fitness graph is altered by flipping the direction of one edge every time  
 the fitness curves of two mutational neighbors intersect (see Fig \ref{fig:intro}{(a) and} (b)). A stable genotype $\vec{\sigma}$ 
 \mm{ceases to be a LFM}  once its fitness curve intersects that of a \mm{mutational} neighbor, and the system transitions to a new stable state by moving along the oriented edges of the new fitness graph. 
 
Given a state $\vec{\sigma}$, we define the 
two disjoint sets 
%\begin{align}
 $I^+[\vec{\sigma}] = \{ i : \sigma_i = 1 \}, 
~ I^-[\vec{\sigma}] = \{ j : \sigma_j = 0\}$. For $i \in I^+[\vec{\sigma}]$, we denote by $\vec{\sigma}^{-i}$ the configuration obtained form $\vec{\sigma}$ by setting $\sigma_i = 0$.  Likewise for $j \in I^-[\vec{\sigma}]$, let 
$\vec{\sigma}^{+j}$ denote the configuration obtained from $\vec{\sigma}$ by setting $\sigma_j = 1$. 
Let  $x_i$ be the intersection point of the dose-response curves of the wild-type $\vec{\sigma} = \vec{0}$ 
and the genotype with a single mutation at site $i$, {\em i.e.} $\vec{0}^{+i}$. Hence,   
$x_i$ is the solution of the equation
\begin{equation}
 w(x) = r_i {w(x/m_i)}.  
 \label{eqn:xdef}
\end{equation}
By a suitable choice of the function $w(x)$ this solution can be guaranteed to be unique (see \cite{das2020predictable} and Appendix B).
% for some $0 < r_i < 1$ and $m_i > 1$. 
It then follows that for $\vec{\sigma}$ and $i \in I^-[\vec{\sigma}]$, the fitness curves of $\vec{\sigma}$ and $\vec{\sigma}^{+i}$ intersect at 
$m_{\vec{\sigma}}  \, x_i$.   
Likewise, {for $j \in I^+[\vec{\sigma}]$} the fitness curves of  $\vec{\sigma}$ and $\vec{\sigma}^{-j}$  intersect at 
$m_{\vec{\sigma}} \, \overline{x}_j$,
where we have defined $\overline{x}_j = \frac{x_j}{m_j}$.  We now see that a necessary and sufficient condition for a genotype $\vec{\sigma}$ to be a stable state is that
% the following inequality holds, 
\begin{equation}
 \max_{j \in I^+[\vec{\sigma}] } \, \overline{x}_j  < \min_{i \in I^-[\vec{\sigma}] } \,  x_i.
 \label{eqn:genPartitionIneq}
\end{equation}
If this holds, let the index $\ell$ {($u$) correspond to the site where the maximum (minimum) on the left (right) hand side of the inequality is  attained}.
% and {similarly, let} the index $u$ {designate  the site where the minimum is achieved} on the right hand side (rhs) of the inequality. 
The stability range of $\vec{\sigma}$ is then 
$(m_{\vec{\sigma}}\overline{x}_\ell, m_{\vec{\sigma}}{x}_u)$. \mm{When starting from a genotype $\vec{\sigma}$} \jk{that is a LFM at $x$},
\mm{the sites $\ell$ and $u$ are the first sites that undergo a mutation when decreasing, respectively, increasing the concentration. We will refer to sites $\ell$ and $u$ as the {\em least-stable} sites.}  

% At this point, we introduce a fruitful analogy with the standard Preisach model, which is comprised of a set of non-interacting
% two-level systems referred to as \textit{hysterons} \cite{preisach1935magnetische,terzi2020state}.
% The mutation variable $\sigma_i \in \{0,1\}$ is analogous to the $i$-th hysteron, and its states $0$ and $1$ to the up and down states of the hysteron. The parameter $x$ plays the role of an external {magnetic} field that drives the system. In the Preisach model, each hysteron has an upper (lower) threshold at which it transitions to the up (down) states as the magnetic field reaches the threshold from the lower (upper) direction. We define the {\it Preisach analogue} of the TIL model as composed of $L$ hysterons, where the upper and lower thresholds of the $i$-th hysteron are $x_i$ and $\overline{x}_i$ respectively. Note that since $m_i>1$, we have $\overline{x}_i < x_i$. There is hysteresis since the transition from the down to the up state occurs at a higher value of the field than the transition from up to down. Referring to Eq. \ref{eqn:genPartitionIneq} we arrive at our first key result: \\
% \textit{The stable states of the TIL model and its Preisach analogue are identical.} \\
% However, as we will show next, there are significant differences in the dynamical properties of the two models.

\mm{At this point, we introduce a fruitful analogy with the standard Preisach model, which is comprised of a set of non-interacting
two-level systems referred to as \textit{hysterons} \cite{preisach1935magnetische,terzi2020state}.
The mutation variable $\sigma_i \in \{0,1\}$ is analogous to the $i$-th hysteron, and its states $0$ and $1$ correspond to the up and down states of the hysteron. The parameter $x$ plays the role of an external {magnetic} field $H$ that drives the system. In the Preisach model, each hysteron has an upper and lower threshold $h^+_i$ and $h^-_i$, respectively. The $i$th hysteron remains in state $0$ as long as $H < h^+_i$, and transits to $1$ otherwise. Likewise for $H> h^-_i$ it remains in state $1$, transitioning to $0$ when this condition does not hold. Imposing for each hysteron that $h^-_i < h^+_i$, implies that in the range $(h^-_i,h^+_i)$ hysteron $i$ can be in either of its two states. The  particular state chosen is history-dependent, giving thereby rise to hysteresis. Thus a necessary and sufficient condition for a hysteron configuration $\vec{\sigma}$ to be attainable at some magnetic field $H$ is that  
\begin{equation}
 \max_{j \in I^+[\vec{\sigma}] } \, h^-_j  < \min_{i \in I^-[\vec{\sigma}] } \,  h^+_i, 
 \label{eqn:genPartitionIneqPR}
\end{equation}
which is identical to the TIL stability condition, Eq.~\eqref{eqn:genPartitionIneq}. We thus define the {\it Preisach analogue} of the TIL model as composed of $L$ hysterons, where the upper and lower thresholds of the $i$-th hysteron are $h^+_i = x_i$ and $h^-_i = \overline{x}_i$, respectively. Note that since $m_i>1$, we have $\overline{x}_i < x_i$, so that the conditions Eq.~\eqref{eqn:genPartitionIneq} and \eqref{eqn:genPartitionIneqPR} are in fact equivalent, and we arrive at our first key result: \\
\textit{The stable states of the TIL model and its Preisach analogue are identical.} }\\

\mm{One immediate result following from the Preisach-TIL equivalence is the following \sgdnew{{\em proximity property}} of stable states which holds for both models: if $\vec{\sigma}$ is a stable state, and \sgdnew{$\ell$ and $u$} 
are its {\em least-stable sites}, then the genotypes $\vec{\sigma}^{+u}$ and $\vec{\sigma}^{-\ell}$, must be stable states as well.
The proof follows by noting that if $\vec{\sigma}$ is stable and hence the inequality Eq.~\eqref{eqn:genPartitionIneq} holds, then by virtue of $\overline{x}_i < x_i$, this inequality must also hold for $\vec{\sigma}^{+u}$ and $\vec{\sigma}^{-\ell}$. The proof of the proximity property is given in } \jk{Appendix B.}

\mm{For the Preisach model the \sgdnew{proximity property} implies the {\em no-avalanche condition} \cite{terzi2020state}: when $H = h^+_u$, we have a transition from $\vec{\sigma}$ to $\vec{\sigma}^{+u}$, such that (i) $\vec{\sigma}^{+u}$ is a stable state by \sgdnew{peak proximity}, and moreover, (ii) $\vec{\sigma}^{+u}$ is a LFM at the field $H = h^+_u$ that triggered the transition and hence no further state changes occur. An analogous result holds when $H = h^-_\ell$. However, while the \sgdnew{proximity property} (i) holds for the TIL model as well, the additional dependence on $m_{\vec{\sigma}}$ of the stability range $(m_{\vec{\sigma}}\overline{x}_\ell, m_{\vec{\sigma}}{x}_u)$ of $\vec{\sigma}$  implies that (ii) will not hold in general. As we will show next, this  leads to significant differences in the dynamical properties of the two models.} \jk{In particular, the TIL dynamics generically includes avalanches.}

\subsection{Dynamics and the transition graph}

\sgd{In this subsection, we discuss dynamics under quasistatically changing $x$. While this assumption is not realistic in clinical applications, it is amenable to analytical treatment and 
provides a stepping stone towards more realistic protocols\mm{, which we will consider in subsection \ref{subsec:beyondQS}}. Crucial aspects of} the dynamics under quasistatically changing $x$ can be described by transitions among stable states\mm{, such that the concentration is changed just enough so that the state ceases to be a LFM}. We call \mm{such} transition\mm{s} under concentration increase a U-transition, and that under concentration decrease a D-transition. Then the dynamics can be described by a {\it transition graph} (see Fig \ref{fig:intro}(c) for an example) where the nodes are the stable states, and each node 
has outgoing U and D edges corresponding to increasing and decreasing concentrations respectively. This must be distinguished from a fitness graph, which is defined at a fixed concentration, and where the \st{directed edges connecting mutational neighbors indicate the direction along which the fitness increases.} 
\begin{figure*}%[\sidecaptionrelwidth][t]
\centering
\includegraphics[height=8cm,width=17.2cm]{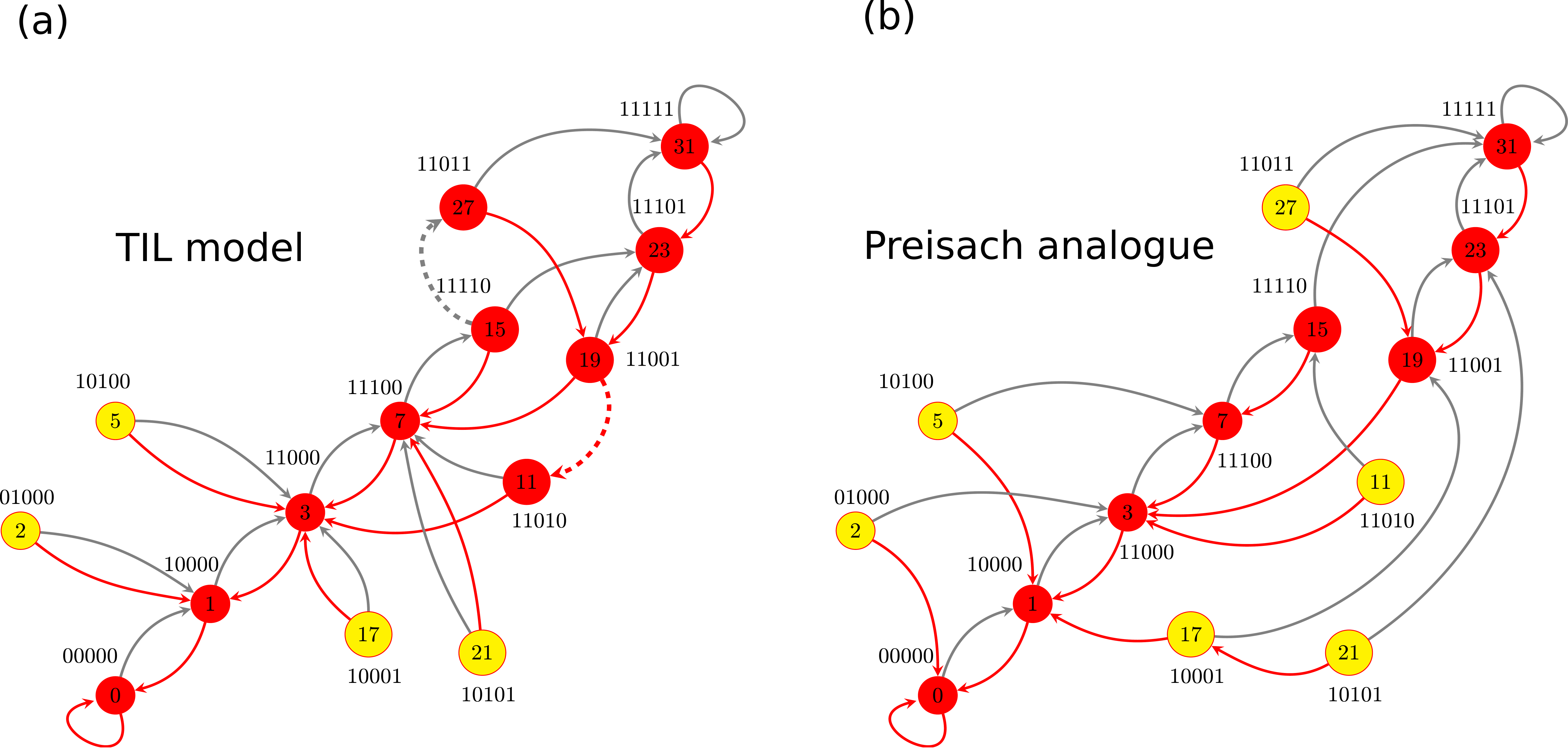}
\caption{\textbf{Transition graph of a realization of the TIL model (left) and its Preisach analogue (right) with $L=5$ sites.}
  %The null-fitness and resistance parameters $(r_i,m_i)$ of the 5 mutations were drawn %randomly from a distribution specified
  %in Appendix B. 
  The symbolic ordering sequence of this realization is given in Eq. \ref{eqn:pexample}. Each genotype is
  assigned an integer label, placed within the nodes, by interpreting the genotype string as a binary code where the leftmost digit is the least significant. The \mm{grey arrows are U-transitions and red arrows are D transitions}. The yellow nodes are the genotypes that cannot be reached starting from the wild type. When multiple outgoing arrows are present from a state, the solid ones correspond to greedy walks, whereas the dashed lines represent fitness-increasing walks but with one or more steps that are not maximally fitness-increasing.}
\label{fig:trgraphs}
\end{figure*}

While the TIL model and its Preisach analogue share the same set of stable states, the dynamical properties are in general different. To illustrate this, in Fig \ref{fig:trgraphs} we show a particular realization of the TIL model with $L=5$ along with its
Preisach analogue.
% We have associated an integer label with each node by using the binary-encoding of the genotype string, as explained in the caption.
In the Preisach model, each transition comprises of a single switching of the least stable hysteron, which leads to a new stable state. In the TIL model, a change at a single site need not lead to a stable state. For example, the U-transition $15 \to 27$ in the TIL model in Fig \ref{fig:trgraphs} involves changes at the third and fifth sites.

To understand why, 
first notice that under concentration changes in the U direction, the first change in a stable state $\vec{\sigma}$ 
(which must satisfy Eq. \ref{eqn:genPartitionIneq}) is a flip $0 \to 1$ at
the site $u$ which has the smallest $x_i$ among sites with $\sigma_i=0$. This flip occurs at 
$x=m_{\vec{\sigma}} x_u$. \mm{By the proximity-property of the TIL model, the new state $\vec{\sigma}^{+u}$ is also stable, satisfying  Eq. \ref{eqn:genPartitionIneq}.}  
%(since $\overline{x_u}< x_u$) and is therefore a stable state. 
However, in order for $\vec{\sigma}^{+u}$ to be a LFM at \mm{$x = m_{\vec{\sigma}} x_u$} we must require that 
its lower stability threshold is less or equal to $x$.
%(equality is allowed, since at $x=m_{\vec{\sigma}} x_u$ the fitness of both $\vec{\sigma}$ and $\vec{\sigma}^{+u}$ are identical, so that 
%the concentration $x$ that triggered the transition must have been slightly above $m_{\vec{\sigma}} x_u$).
% whether it is a local fitness maximum at $x$ depends on whether the lower 
% stability threshold of $\vec{\sigma}^{+u}$ is smaller than $x$.  
This threshold is  
$m_{\vec{\sigma}^{+u}} \overline{x}_j = m_{\vec{\sigma}} m_u \overline{x}_j$,  
where {$j\in I^+[\vec{\sigma}^{+u}]$} and $\overline{x}_j > \overline{x}_k$ for all {$k \ne j$ with $k\in I^+[\vec{\sigma}^{+u}]$}. Therefore 
the new state is a fitness maximum if and only if $m_u \frac{\overline{x}_j}{x_u} = {\frac{\overline{x}_j}{\overline{x}_u}\le 1}$, in which case the dynamics terminates at $\vec{\sigma}^{+u}$. When the condition is violated, additional {\it secondary mutations} occur until a fitness maximum is reached. \jk{In the terminology of driven disordered systems, this corresponds to the occurrence of an avalanche}. 
While many detailed properties of the secondary mutations
depend on system parameters, certain {features are general and in particular do not depend on the choice of $w(x)$. In the following we will mention 
some of these.}
% general features will be mentioned here \mm{that do not . 

In the Preisach model, exactly $L$ U-transitions are required to
get from the wild type to the all mutant, and exactly $L$ D-transitions to go from the all-mutant to the wild type, as is seen in Fig \ref{fig:trgraphs}. In the TIL model, due to the existence of secondary mutations, these numbers are generally different. The TIL model in Fig \ref{fig:trgraphs} requires $6$ U-transitions to go from the wild type to the all-mutant, and $6$ D-transitions in the reverse direction, even though $L=5$. One important consequence of the secondary mutations is that
\textit{the number of mutations does not always increase monotonically under U or decrease monotonically under D}. The first secondary mutation is always of a complementary kind to the initial mutation, where the changes $0 \to 1$ and $1\to 0$ are defined to be of complementary kind to each other (see {\it Appendix B} for a proof). Further mutations may also continue to be complementary to the initial mutation, leading to a (temporary) decrease in the number of mutations under U or an increase under D, as shown in a typical trajectory for $L=20$ mutations in Fig \ref{fig:UDb}(a). This seems counterintuitive, but arises from the state-dependent pre-factor $m_{\vec{\sigma}}$ in the stability thresholds of stable states.

Moreover, when secondary mutations are present, the state $\vec{\sigma}^\prime$ to which a transition occurs from a state $\vec{\sigma}$ need not be unique, due to the possible presence of multiple adaptive paths. In \st{the TIL graph of} Fig \ref{fig:trgraphs}, the state $15$ can  transition either to the state $27$ or the state $23$ under concentration increase.
It can also be shown, using the property of direct path accessibility, that secondary mutations cannot cause a transition to a subset or superset of $\vec{\sigma}$ (see Appendix B), i.e. both the initial and the final state must contain at least one mutation not contained in the other. Another related consequence of the secondary mutations is that $\vec{\sigma}$ may transition to the same state $\vec{\sigma}^\prime$ under U and D-transitions. For example, the state $5$ in the TIL graph in Fig \ref{fig:trgraphs} transitions to the state $3$ under both U and D-transitions. This also appears counterintuitive from a biological standpoint, but it can occur when the stability range of $\vec{\sigma}$ is contained in that of $\vec{\sigma}^\prime$.
% This implies that  $\vec{\sigma}$ and $\vec{\sigma}^\prime$ cannot be mutational neighbors, and therefore $\vec{\sigma} \to \vec{\sigma}^\prime$ must involve secondary mutations; this further implies that $\vec{\sigma}^\prime$ cannot be a subset or superset of $\vec{\sigma}$. 

%\subsection{Ordering sequences}
\begin{figure}%[tbhp]
\centering
\includegraphics[width=.7\linewidth,height=.15\linewidth]{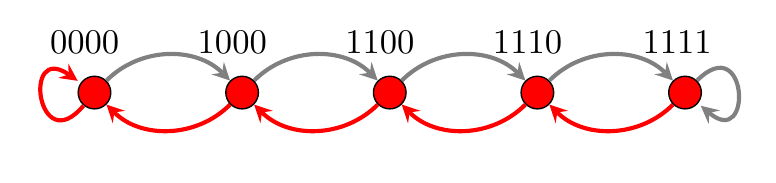}
\caption{\textbf{TIL transition graph with $L=4$ and no secondary mutations.} The ordering sequence is of the form given in Eq. \ref{eqn:pnosecndmut}. The transition graph is unique for this ordering sequence, and is identical to the graph for the Preisach analogue.}
\label{fig:nosecmut}
\end{figure}

To understand the transition graph of the TIL model in a more systematic way, we adopt a strategy that has been fruitful for the Preisach model \cite{terzi2020state}. { We construct a symbolic sequence $p$ that specifies the total order among all the elements of $\{x_i \}_{1}^L$, $\{\overline{x}_j\}_{1}^L$. First, without loss of generality, we order our indices $i$ such that $x_1 < x_2 < \dots < x_L$. Next, it is useful to define the 
permutation $\rho$ of $(1, 2, \ldots, L)$ that orders the $\bar{x}_i$ among themselves from largest to smallest, so that
$\bar{x}_{\rho_1} > \bar{x}_{\rho_2} > \ldots > \bar{x}_{\rho_L}$.}
Since $\overline{x}_i < x_i$ for each $i$, what remains is the specification of the ordering relation between $\overline{x}_i$ and $x_j$ for $j \ne i$. Given the sets $\{x_i \}_{1}^L$, $\{\overline{x}_j\}_{1}^L$ and the ordering prescribed by  
$\rho$, we can describe the total ordering in terms of a symbolic sequence $p$ of elements $\bar{i}$ and $i$ by making the correspondence $\bar{i} \leftrightarrow \bar{x}_i$ and $i \leftrightarrow x_i$,
so that the sequence specifies the increasing order of $x_i$ and $\bar{x}_i$. Since $\bar{x}_i < x_i$ and the permutation $\rho$  have 
to be respected, the sequence has to be such that the following hold: 
for each $i$, $\bar{i}$ is to the left of $i$; the subsequence of sites without overbars is $1, 2, \ldots, L$;
the subsequence of sites with overbars is $\rho_L, \rho_{L-1}, \ldots, \rho_1$. 
As an example, consider the TIL Model in Fig \ref{fig:trgraphs}, which has $L = 5$, $\rho = (43521)$ and the ordering  
\begin{equation}
\overline{x_1} < \overline{x_2} < x_1 < \overline{x_5} < \overline{x_3} < x_2 < \overline{x_4} <  x_3  < x_4  < x_5.
%\label{eqn:porder}
\nonumber
\end{equation}
The corresponding symbolic ordering sequence $p$ is then 
\begin{equation}
p = \quad \overline{1} \quad  \overline{2} \quad 1 \quad  \overline{5} \quad \overline{3} \quad 2 \quad \overline{4} \quad 3 \quad 4 \quad 5.
 \label{eqn:pexample}
\end{equation}
%In general, given $L$, it is readily shown that there are $ N_L = 1\cdot3 \cdot5\cdots (2L - 1) = (2L - 1)!!$ distinct %possible ordering sequences (see Mathematical Methods). 
%The mathematical details and proofs are to be found in the Mathematical Methods section. 
% For example, whether a genotype is stable or not, and what the least stable sites of a stable state are, can be read off easily from $p$, since Eq. \ref{eqn:genPartitionIneq} is easy to check by inspecting $p$. 

%\subsection{Ordering sequences}
\begin{figure}%[tbhp]
\centering
\includegraphics[width=\columnwidth]{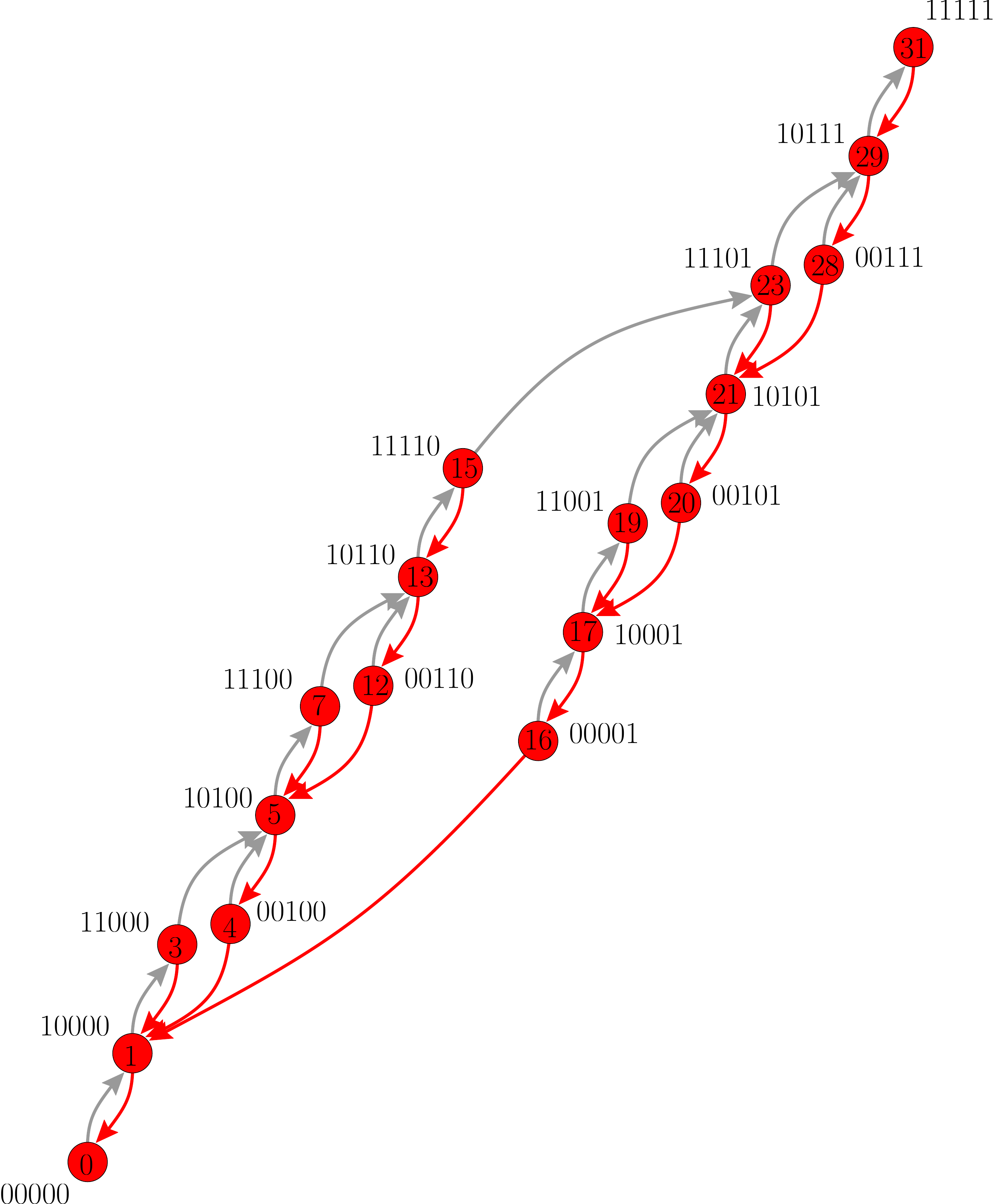}
\caption{\textbf{TIL transition graph with $L=5$ and nested hysteresis loops.} \sgd{This graph was generated from a system with tradeoff among all pairs of mutations, {\em i.e} \mm{requiring that} $r_i < r_j \iff m_i > m_j$ for all $i,j$, \mm{and considering only transitions under greedy adaptive walks}.} \mm{The nesting of graph loops implies that genotypes can partially encode past changes of concentration. For example, starting from the wildtype $0$ at concentration $x = 0$, the genotype $19$ can be realized only by increasing first the concentration enough in order to reach at least $23$, followed by a decrease leading to at least $17$ (but not further than $16$) and a final increase of concentration. }
}
\label{fig:TILtnested}
\end{figure}

Because of the Preisach-TIL correspondence, whether a genotype/Preisach state is stable or not, and what the least stable sites of a stable state are, can be read off from $p$, since the condition in Eq. \ref{eqn:genPartitionIneq} is easy to check by inspecting $p$ \mm{(details are given in Appendix B)}.
In the case of the Preisach model this implies that $p$ completely determines the transition graph \cite{terzi2020state}. While this is not the case for the TIL model, the sequence $p$ nevertheless contains considerable information about the 
TIL transition graph. In particular, this representation provides a precise condition for the existence of secondary mutations
(see Appendix B): \\
\textit{Secondary mutations are absent from all transitions in the TIL model if and only if the ordering sequence is of the form}
    \begin{equation}
     p = \overline{1} \quad 1 \quad \overline{2} \quad 2 \quad \dots \quad \overline{L} \quad L.
     \label{eqn:pnosecndmut}
    \end{equation}
    In the absence of secondary mutations, the transition graph of the TIL model becomes identical to that of its Preisach analogue. 
    The transition graph in this case has a simple chain structure (see Fig \ref{fig:nosecmut}), and the number of stable states
    is $L+1$, which is the lowest possible in a TIL (or Preisach) model. Note that despite identical transition graphs in this case, some dynamical differences are still present. Each Preisach element is hysteretic, and therefore forward and reverse transitions between two states do not occur at the same concentration; in the TIL model satisfying Eq. \ref{eqn:pnosecndmut}, however, they occur at the same concentration, namely the one at which the dose-response curves of the two genotypes intersect.         

\subsection{Hysteresis, reversibility and memory} 
The reversibility of evolution under a reversal of environmental conditions is an important question in evolutionary biology
\cite{teotonio2001reverse,bridgham2009epistatic,kaltenbach2015reverse}.
In the specific case of antibiotic resistance evolution, to what extent resistance is reversed in a drug-free environment is a question of considerable clinical importance \cite{andersson2010antibiotic,allen2017reversing,durao2018evolutionary}. One should note that different notions of reversion are used here. One common definition refers to a sudden (rather than slow) change in environment to a new state, followed by a switch back to the original state \cite{andersson2010antibiotic,dunai2019rapid,zur2010compensation}. 
In the context of our model, we \mm{consider first} reversion under quasistatic environmental changes, which would appear to be most conducive
for approximately reversible behavior. The phenomenon \mm{of reversion} is naturally linked to the notion of hysteresis under a slow and continuous change of an external field, and indeed the Preisach model was \mm{originally} proposed as a simplified, tractable model of hysteresis \st{in magnetic materials} 
\cite{preisach1935magnetische,terzi2020state}. \mm{We then turn in section \ref{subsec:beyondQS} to the possibility of reversion under jump-like concentration changes.}

\mm{The TIL model} also exhibits hysteresis and irreversibility.
% ) as shown in Fig \ref{fig:trgraphs}.
The highest degree of reversibility is exhibited by systems with chain-like transition graphs, such as in Fig \ref{fig:intro}(c) or Fig \ref{fig:nosecmut}, where each transition $\vec{\sigma} \to \vec{\sigma}^\prime$ is accompanied by the transition $\vec{\sigma}^\prime \to \vec{\sigma}$ in the reverse direction, and there {are no states with} multiple outgoing edges in either direction. This means that under a reversal of the direction of concentration change, the same genotypes occur in reversed sequence. However, the transitions $\vec{\sigma} \to \vec{\sigma}^\prime$ and $\vec{\sigma}^\prime \to \vec{\sigma}$ need not occur at the same concentration. For example, in Fig \ref{fig:intro}(d), the transition $10 \to 01$ occurs at the point $x=c$ during concentration increase, but $01 \to 10$ occurs at $x=b$ during concentration decrease. On the other hand, for systems of the type shown in Fig \ref{fig:nosecmut}, the forward and reverse transitions occur at the same concentration. 
However, such perfect reversibility is not typical of TIL models. \sgd{The degree of reversibility depends on parameter choices.} \jk{Figure}~\ref{fig:TILtnested} shows a realization of the TIL model with $L=5$ loci and a high-degree of irreversibility, i.e. forward transitions with no corresponding reverse transitions, such as the \mm{D-transition  $16 \to 1$ or the U-transition $15 \to 23$}.
% in the TIL graph in Fig \ref{fig:TILtnested}. 

Based on the observation of the systems described in Fig \ref{fig:intro} and \ref{fig:nosecmut}, we need to distinguish between two kinds of hysteresis loops. We say that two states $\vec{\sigma}$ and $\vec{\sigma^\prime}$ form a { \it concentration loop} if  one can go from $\vec{\sigma}$ to $\vec{\sigma^\prime}$ under quasistatic concentration increase and from  $\vec{\sigma^\prime}$ to $\vec{\sigma}$ under concentration decrease, { and there is some range of $x$ over which the forward and reverse trajectories do not share the same genotype}. The system in Fig \ref{fig:intro} exhibits a concentration loop, as shown by the rectangle with corners marked by the points $p,q,r$, and $s$  in Fig \ref{fig:intro}(d).
We say that $\vec{\sigma}$ and $\vec{\sigma}^\prime$ form a {\it graph loop} $(\vec{\sigma}$,$\vec{\sigma}^\prime)$ if one can go from $\vec{\sigma}$ to $\vec{\sigma^\prime}$
under U-transitions and from $\vec{\sigma}$ to $\vec{\sigma^\prime}$ under D-transitions,  and 
if there is at least one genotype contained in either the forward or reversed sequence of states that is not contained  in the other. A graph loop between two states implies a concentration loop, but a concentration loop does not imply a graph loop. For example, the case shown in Fig \ref{fig:intro} has a concentration loop, but does not have a graph loop, {as can be seen} in Fig \ref{fig:intro}(c).  An example of a graph loop in the TIL model in Fig \ref{fig:trgraphs}(a) is the one formed by the \mm{sequence of U-transitions from $1$ to $23$ and the D-transitions leading from $23$ back to $1$.} A necessary condition for the existence of graph loops %({$\vec{\sigma}$,$\vec{\sigma}^\prime$) 
in the TIL model is the presence of secondary mutations; 
for otherwise, every transition must be among mutational neighbors, and such that the upper stability threshold of one coincides with the lower stability threshold of the other, causing the transitions to be reversible.

Hysteresis is also linked to the notion of memory \cite{keim2019memory,mungan2019networks}. A genotype encountered along a trajectory not only
contains information about the concentration, but also about the history of environmental change. At the simplest level, it may contain information about whether one is on the U or D boundary of {a} 
loop. For example, in the region between $x=b$ and $x=c$ in Fig \ref{fig:intro}(d), the state $10$ indicates that the concentration has been {increasing,}
%on an upward trajectory, and 
{while $01$ indicates that it was decreasing.}
%the downward trajectory. 
But there is more information available than this, in general. The subloops seen in the {TIL}
%Preisach 
graphs in \mm{Figs. \ref{fig:trgraphs} and \ref{fig:TILtnested}} contain (partial) information about extreme values of $x$ reached in previous rounds of concentration cycling. 
%NEW TEXT:
\mm{For example in the graph shown in Fig.~\ref{fig:TILtnested}, if the dynamics started with the wild type at $x=0$ and reached the genotype $19$ at some point, one infers that the last transition happened by increasing the concentration to above the stability threshold of $17$; but we also see from the transition graph that on some previous upward path the concentration must have exceeded the upper stability threshold of $15$, followed by some sequence of transitions that brought it to the lower stability threshold of $16$ for the first time since this happened.}
% %OLD TEXT
% Thus, if the dynamics started with the wild type at $x=0$ and reached the genotype $11$ at some point, one infers that the last transition happened by a lowering of the concentration to below the stability threshold of $19$; but we also see from the transition graph that on some previous upward path the concentration must have exceeded the upper stability threshold of $15$, followed by some sequence of transitions that brought it to the lower stability threshold of $19$ for the first time since this happened.

In this context, it is important to mention the phenomenon of {\it return point memory} (RPM) 
\cite{sethna1993hysteresis,munganterzi2018} possessed by certain \st{ systems. In}    
  our setting of adaptive evolution, \st{RPM} implies that genotypes at which the direction of the concentration
  change has been reversed can be returned to with a subsequent reversal and hence remembered. The RPM property is universally present in the Preisach model \cite{keim2019memory,munganterzi2018,terzi2020state}. In the context of state transition graphs, one talks about the loop-RPM property, which ensures that the system cannot escape any loop between two states $\vec{\sigma}$ and $\vec{\sigma^\prime}$ without passing through one of these states (see \cite{munganterzi2018,terzi2020state} for a detailed exposition). The RPM property implies the loop-RPM property, and is therefore possessed by the Preisach model and can be checked
  for the Preisach {{graph}} in Fig \ref{fig:trgraphs}. 
% and also TIL transition graphs in Fig \ref{fig:trgraphs}.

% NEW TEXT
Return-point memory is a mechanism by which a memory of local extremes of the driving parameter can be retained. For example, in the TIL model in Fig \ref{fig:trgraphs}(a) and under greedy dynamics, starting with the wild type at $x=0$, and increasing the concentration until state $15$ is reached, any decrease of concentration followed by a subsequent increase will eventually lead again to state $15$. However, if the concentration continues to increase, so that state $23$ is reached, then a concentration decrease to \mm{say $19$} followed by an increase will not lead to $15$ anymore. Thus the memory of $15$ as the genotype at a local extreme event has been erased and replaced by $23$. While we have found many realizations of the TIL model that possess the loop-RPM property under greedy transitions, \mm{such as the transition graphs shown in Figs.~\ref{fig:trgraphs} and \ref{fig:TILtnested}}, it is not universally present. Additionally, the existence of alternative fitness increasing transitions, such as the ones shown in 
Fig \ref{fig:trgraphs} by the dashed lines, can cause a loss of this kind of memory. To see this,  consider the loop formed by the greedy U transitions leading from state $7$ to $23$ and the greedy D transitions going from  $23$ to state $7$. On the downward trajectory from $23$, it is possible to escape the loop without going first through $7$ by making the transition $19 \to 11$.

\subsection{\jk{Expected} number of stable states}

\begin{figure*}%[\captionrelwidth][t]
\centering
\includegraphics[height=7cm,width=17cm]{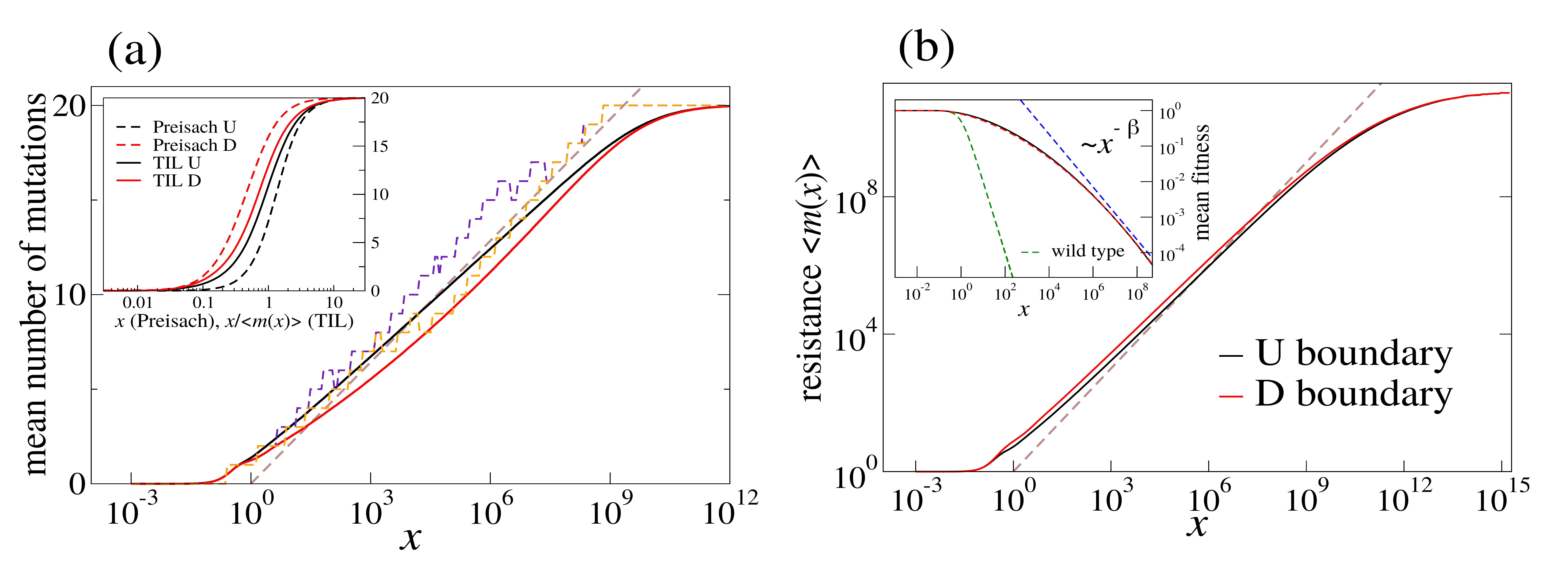}
\caption{\textbf{Average properties of evolutionary trajectories along the main loop for $L=20$.}
  The dose-response curve is $w(x)=1/(1+x^2)$. The mutation parameters $(r_i, m_i)$ are drawn randomly
  from a joint probability density described by Eqs. \eqref{eq:dist1} and \eqref{eq:dist2} in Appendix C. Adaptive walks \sgdnew{(based on selection coefficients) were used for evolutionary dynamics. Greedy walks produce very similar results and are not shown}. A total of $10^4$ realizations were used to calculate averages. ({\bf a}) Mean number of mutations in the genotypes along the U and D boundaries, which describe the behavior under increasing (U) and decreasing (D) concentrations. The dashed purple and orange lines show a typical sample trajectory. The dashed brown line is the curve $\frac{\ln x}{b}$, where 
  $b=\langle \ln m \rangle$.
  % The expectation value is with respect to $P(r,m)$.
  The inset shows rescaled values of the mutation number in comparison to the Preisach analogue (dashed lines). The rescaling is done by the mean resistance level $\langle m(x) \rangle$.
  % where $m(x)$ is the resistance level of the genotype at $x$, and average is with respect to the density $Q(\cdot)$.
  ({\bf b}) Mean resistance level $\langle m(x) \rangle$ of the genotypes along the boundaries. The brown dashed curve is $x$. The inset shows the fitness of the genotypes along the boundaries. The dashed green line is the fitness of the wild type $w(x)$, and the
  dashed blue line is the power law $x^{-\beta}$, where $\beta = -\frac{a}{b}$ and $a=-\langle \ln r \rangle$. 
  Angular brackets denote averages with respect to the distribution $Q(\{(r_i,m_i)\})$ (see main text for further details).
  }
\label{fig:UDb}
\end{figure*}

While \st{many properties of the TIL model depend only on the ordering sequences $\rho$ and $p$,} 
% encoded by the transition
% graph, 
for a more detailed study of the concentration-dependent evolutionary dynamics the fitness values of the model need to
be explicitly assigned. In the standard Preisach model, one usually considers the thresholds to be independent random variables. 
Similarly for the TIL model, we assume that $r_i$ and $m_i$ follow a joint {probability distribution with density $P(r,m)$
and the ordered pairs $(r_i,m_i)$ {for $i = 1, 2, \ldots L$ are independently and identically distributed}. Their joint probability density is then given by  $Q(\{(r_i,m_i)\})=\prod_{j=1}^L P(r_j,m_j)$. 
%For this distribution, $x_i$ is in general correlated with $\bar{x}_i$, but  $x_j$ and $\overline{x}_j$ for $j \ne i%%$. 

Since $x_i$ and $\bar{x}_i$ are functions of $r_i$ and $m_i$ only, the pairs $(x_i,\bar{x}_i)$ for $i = 1, 2, \ldots L$ are independently and identically distributed as well. 
Let the (marginal) cumulative distribution function (CDF) of $x_i$ be $F_x (x_i)$ and that of $\bar{x}_i$ be 
$F_{\bar{x}}(\overline{x}_i)$, and let $P_{\bar{x}}(z)=F_{\bar{x}}^\prime(z)$ {denote the} probability density function of $\overline{x}_i$.
 The probability that a genotype is a fitness maximum is the probability that Eq. \ref{eqn:genPartitionIneq} holds. The calculation of this probability is facilitated by the fact that $I^+[\vec{\sigma}]$ and $I^-[\vec{\sigma}]$ are disjoint sets. One can show that in the limit of large $L$ 
 the average number of stable states $\langle  N_{ss} \rangle$ is given by
 \begin{equation}
%   \overline{N_{ss}} \simeq \sqrt{\frac{{2 \pi}}{{(L-1) | G^{\prime \prime}(z_0) |}}} 2^{\frac{G(z_0)}{\ln 2} (L-1)}, 
  \langle {N_{ss}} \rangle \simeq \sqrt{\frac{{2 \pi}}{{(L-1) | G^{\prime \prime}(z_0) |}}} {\, {\rm e}^{G(z_0) (L-1)}}, 
  \label{eq:nss}
 \end{equation}
where the average is taken with
% $N_{ss}$ is the number of stable states, and angular brackets $\langle \cdot \rangle$ denote average with 
respect to $Q(\{r_i,m_i\})$, 
$G(z)=\ln (1+F_{\bar{x}}(z)-F_{{x}}(z))$, and the global maximum of this function is at $z_0$. Thus the mean number of stable states is asymptotically exponential in $L$, showing the highly rugged nature of these landscapes
(see Appendix C for the derivation of Eq. \ref{eq:nss}).

\begin{figure*}
\centering
\includegraphics[width=18cm,height=7cm]{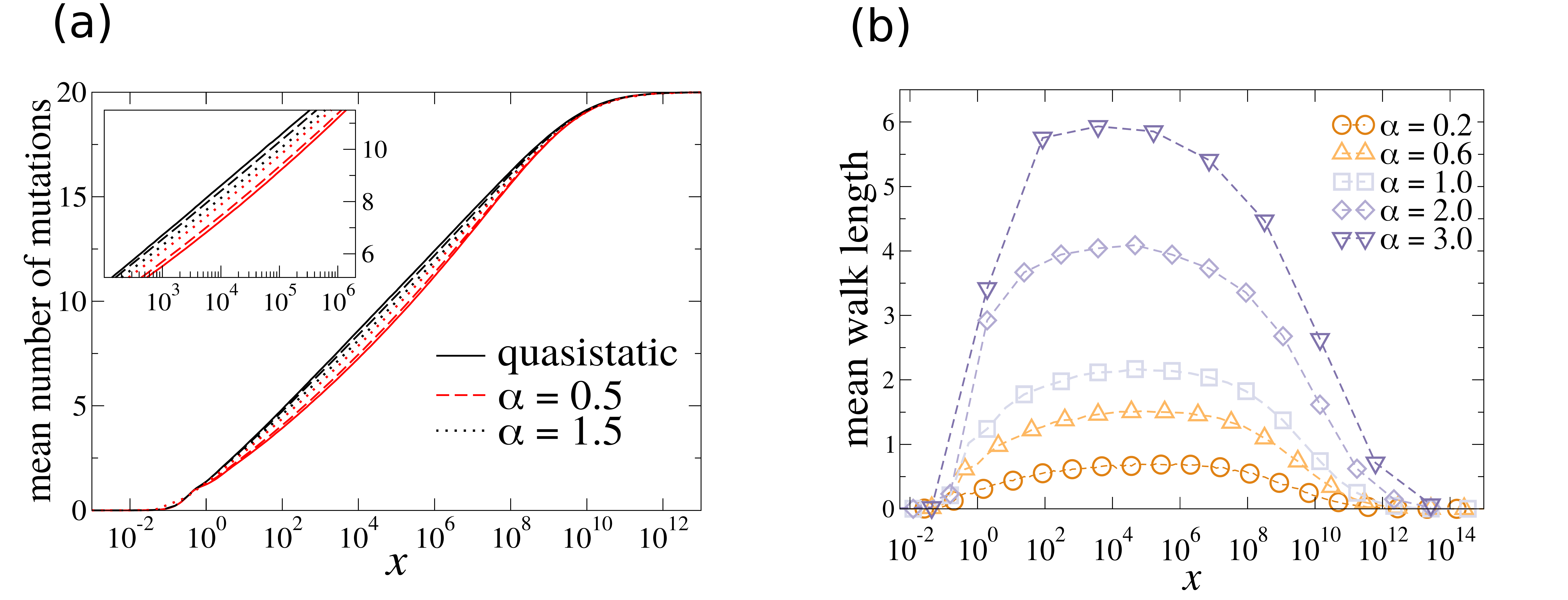}
\caption{\sgd{\bf Average properties of evolutionary trajectories along the main loop under finite driving rates $\alpha$}. (a) Simulation results for the mean number of mutations along the main hysteresis loops have been plotted for different values of the jump sizes in concentration. The concentration has been changed in discrete steps by factor of $\langle m \rangle ^\alpha$. The quasistatic case corresponds to the limit $\alpha \to 0$. The inset shows a zoomed-in version for clarity. (b) The length of adaptive walks has been plotted as a function of concentration for various values of $\alpha$. \jk{Adaptive walk lengths exceeding one step imply that
secondary mutations have occurred.}}
%\textbf
%{
\iffalse
Mean Hamming distance between genotypes on the U and D boundary as a function of concentration for L=20.} A total of $10^5$ realizations were used to calculate averages.
  The distance measure $d_U(x)$ ($d_D(x)$) is the minimal Hamming distance between the current genotype on the U (D) boundary and
  any point on the D (U) boundary. A low value corresponds to high reversibility. The black (red) dotted line is the concentration at which the mean number of mutations is $L/2$ on the U (D) boundary, obtained from Fig \ref{fig:UDb}(a).
  \fi
 % }
\label{fig:jumps}
\end{figure*}

\subsection{\mm{Reversibility in the main hysteresis loop}}
\label{subsec:reversibility}

An important dynamical question is understanding the {evolutionary sequence of genotypes}
%trajectory of the system 
as the concentration is cycled between very low and very high values. 
\sgd{We numerically generate trajectories starting from the wild type at $x=0$ and increasing $x$ until the all-mutant is reached, and then decreasing $x$ until the wild type is reached again. At each $x$, the system is evolved through the adaptive walk \sgdnew{(based on selection coefficients, as discussed in Section II)} \mm{until} an LFM is reached.
 We call the trajectory comprised of the sequences of LFMs the {\it main {hysteresis} loop}, and the upward and downward parts of it the U- and D-boundary respectively.}
The mean number of mutations in the genotype under quasistatic change in $x$ on both the U and D boundaries are shown in Fig \ref{fig:UDb}(a), which clearly shows hysteresis.
% (see caption of Fig \ref{fig:UDb} for details of the distribution used).
The inset shows a comparison with the Preisach model. \mm{Recall} that in the TIL model the range of concentrations over which a genotype $\vec{\sigma}$ is a local maximum has an overall scale factor $m_{\vec{\sigma}}$. Therefore, in order to facilitate comparison of the data for the TIL model and its Preisach analogue, we have rescaled the concentration axis for the former by $\langle m(x) \rangle$, 
the average scale factor $m_{\vec{\sigma}}$ {of the states $\vec{\sigma}$ that are stable} at concentration $x$.
% Here the $x$ values for the TIL model have been scaled by the $\langle m(x) \rangle$ (where $m(x)$ is equal to $m_{\vec{\sigma}}$ of the stable genotype at $x$) to facilitate the comparison.
{From the inset of \ref{fig:UDb}(a) we see that for the Preisach model} the curve for the number of mutations on the U-boundary \mm{lies below}  the corresponding curve for the D-boundary. This effect can be understood qualitatively as follows. The intersection points along the U-boundary are governed by the distribution of $x_i$ and those along the D-boundary by the distribution of $\bar{x}_i = x_i / m_i < x_i$. As a result, the $i$-th mutation is acquired at a larger $x$ along the U-boundary compared to where it is lost on the D boundary. 
More generally, for any randomly chosen pair of mutations $i$ and $j$, $x_j$ tends to be higher than $\bar{x}_i$ since all the $m_i$'s are larger than $1$. The consequence is that the intersections along the U-boundary tend to occur at larger values of $x$ compared to the D-boundary, making the curve for the number of mutations along the U-boundary lower. Essentially the same effect is visible for the TIL model when the $x$ values are rescaled by the value of $\langle m(x) \rangle$ on the boundaries.
When the rescaling is not done, the U-boundary becomes higher, as seen in {the main} Fig \ref{fig:UDb}(a).
\begin{figure*}%[\sidecaptionrelwidth][t]
\centering
\includegraphics[width=.92\linewidth]{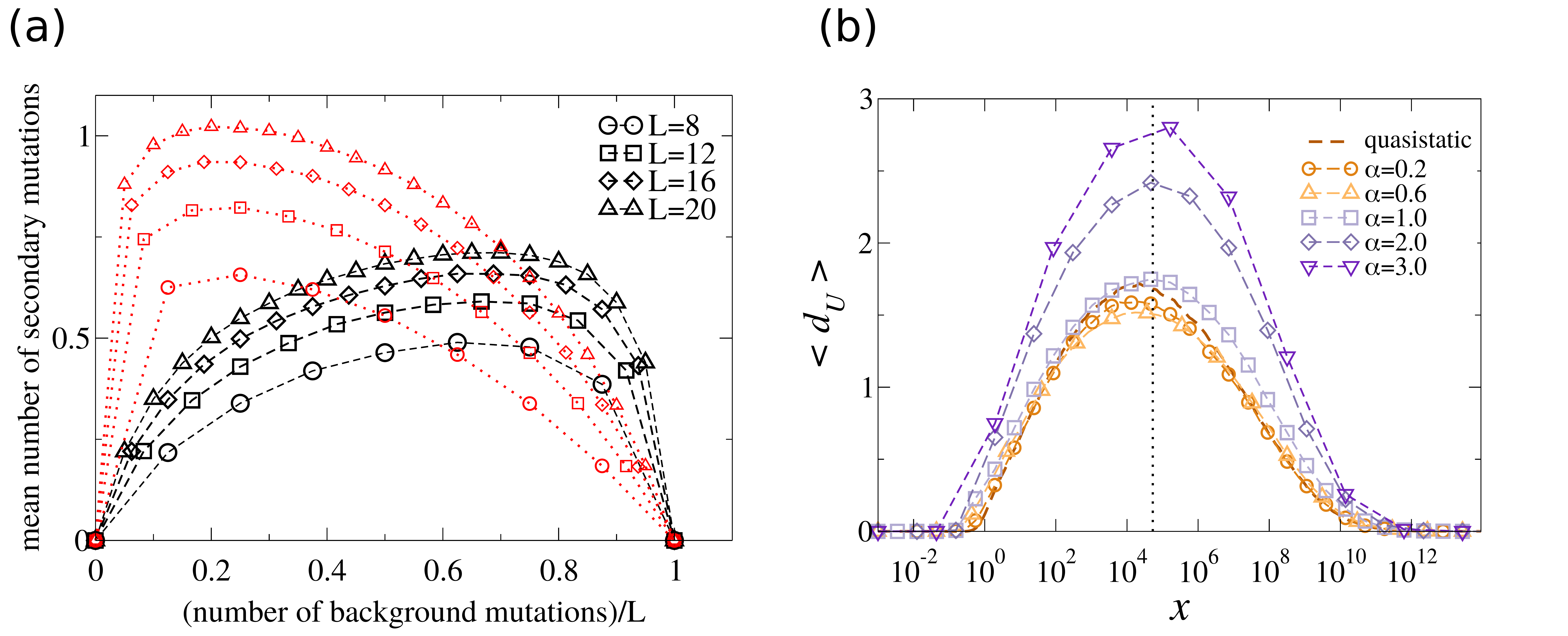}
\caption{\textbf{Results on secondary mutations and genotypic reversibility}. \sgd{Simulations were conducted along the main {hysteresis} loop, and the mutation parameters
  were generated randomly as described in the text and Appendix C. (a) This figure shows the mean number of secondary mutations in an evolutionary transition \sgd{under quasistatic driving} as a function of the number of mutations
    in the originating background genotype. We used $10^5$ realizations for averaging. Black symbols correspond to the U boundary and red to the D boundary. 
  (b) Mean path irreversibility $\langle d_U(x) \rangle$ is shown for various values of the driving rate $\alpha$. Averages were performed over $10^4$ realizations. The dashed black line corresponds to the quasistatic limit $\alpha \to 0$.}      
  }
\label{fig:irrev}
\end{figure*}
The clue to understanding this {comes}
%is obtained 
from Fig \ref{fig:UDb}(b), which shows that the average resistance level $\langle m(x) \rangle$ at given $x$ is lower for the U-boundary. Since the intersection points have the pre-factor $m_{\vec{\sigma}}$ in the TIL model, this effect tends to make the intersection points along the U-boundary occur at lower values of $x$. For our system, this effect is 
{apparently} 
strong enough to shift the curve of the number of mutations along the U boundary above that of the D-boundary. 

Generally, the changes of resistance level and mutation number have a complex mutual dependence, and these can vary between systems depending on the dose response curve and the parameter distribution. However, certain asymptotic features that hold generically for stable states can be computed to \mm{leading order.} For example, as shown in Appendix C, the mean number of mutations in a state that is stable at $x$ scales asymptotically as 
$\simeq \frac{\ln x}{b}$, {where the parameter $b=\langle {\ln m} \rangle$, and $m$ is the resistance of an individual mutation.} This is indicated by the
%shown as a 
brown dashed line in Fig \ref{fig:UDb}(a). %, which shows that $L=20$ the asymptotic result already provides a reasonably good approximation in the range $1 \ll \ln x/b \ll L$.
%MM: replaced the previous text with this: (provided $1 \ll \ln x/b \ll L$).  
At the same level of approximation, the mean level of resistance satisfies the relation $\ln \langle m(x) \rangle \simeq \ln x$,
which is shown as a dashed brown line in Fig \ref{fig:UDb}(b). The inset of  Fig \ref{fig:UDb}(b) displays the fitness as a function of $x$, which is seen to decline at a much lower rate than that of the wild type, as a consequence of the increasing level of
resistance. Detailed derivations of these results are given in Appendix C.
%\vspace{65pt}

\subsection{\mm{Beyond the quasistatic approximation}}
\label{subsec:beyondQS}

\sgd{For many applications, a more realistic driving protocol is \mm{one} where $x$ is changed \mm{discontinuously}. Experiments on antibiotic resistance evolution often consider protocols where the concentration is changed stepwise by a constant factor \cite{baym2016spatiotemporal,schenk2022population}. To simulate this scenario, we numerically implemented a driving protocol where concentration \mm{is} sequentially increased in discrete steps from very low to very high values and then reversed. Both the increase and decrease \mm{occur} by a factor of $\langle m \rangle^\alpha$ in each step, and several value\mm{s} of $\alpha$ were used. The limit $\alpha \to 0$ corresponds to quasistatic driving. The results for the mean number of mutations along the U- and D-boundaries are plotted in Fig \ref{fig:jumps}(a). The hysteresis loops are thinner, implying a higher phenotypic reversibility. When the concentration changes by a finite amount, the subsequent adaptive walk can move stochastically along multiple paths involving larger numbers of mutations (see Fig \ref{fig:jumps}(b)). One therefore expects the system to reach a final distribution of phenotypes that depends less strongly on the starting point, consistent with higher reversibility on the phenotypic level (but not necessarily on the genotypic level, as we\mm{ will} see in the next subsection).      
}

\subsection{\sgd{Genotypic reversibility}} 
Figure \ref{fig:irrev}(a) shows simulation results \sgd{with quasistatically changing $x$} for 
the mean number of secondary mutations as a function of the number of background mutations (i.e., the number of mutations in the genotype from which the transition originates) along the main {hysteresis} loop. For large $L$, the number of secondary mutations depends weakly on the number of background mutations (unless the latter is close to $0$ or $L$).
Moreover, as $L$ increases linearly, the gaps between the curves decrease, indicating a possible asymptotic convergence towards a limiting shape (although this could not be verified conclusively).
The number of secondary mutations is seen to be small, implying that adaptive walks at fixed $x$
\jk{are} short for quasistatic changes in $x$. 

As was shown above, secondary mutations are also the source of {{genotypic}} irreversibility under quasistatic drive in the TIL model. We now describe a measure of genotypic irreversibility, adapted from a distance measure for evolutionary paths introduced in \cite{lobkovsky2012replaying}. Let ${\vec{\sigma}}$ be the genotype at $x$ on the U boundary of the main loop. Then $d_U (x)$ is defined as the minimum of the Hamming distance between ${\vec{\sigma}}$ and the genotypes on the D boundary (for any \jk{concentration}). 
The quantity $d_D(x)$ can be defined in an analogous way. The quantity $\langle d_U(x) \rangle$ is plotted in Fig \ref{fig:irrev}(b). The distance measures vanish at very low and high concentrations, which is expected since the wild type and all-mutant are {on both the U and D boundaries of the loop.} 
% reached by both U and D trajectories. 
The maximum value is reached close to the concentration at which the estimated mean number of mutations is $L/2$, as shown by the vertical dotted line.  The level of
reversibility is quite high for quasistatic driving, consistent with the low number of secondary mutations seen in Fig \ref{fig:irrev}(a). \sgd{The quantity $\langle d_D(x) \rangle$ shows very similar behavior (not shown).} 

\sgd{We have also quantified the genotypic irreversibility for finite $\alpha$ using the measure $\langle d_U(x) \rangle$, which is shown in Fig \ref{fig:irrev}(b). For large $\alpha$, the genotypic reversibility is lower, in contrast to the phenotypic reversibility in Fig \ref{fig:jumps}(a). The walk lengths increase with increasing $\alpha$ as shown in Fig \ref{fig:jumps}(b). The longer walks lead to higher genotypic divergence, even though the phenotypic properties of the LFMs that are accessed are very similar, as demonstrated in Fig \ref{fig:jumps}\mm{(a)}. Thus, even when there is substantial hysteresis at the genotypic level, one may still observe very similar levels of drug resistance evolving along the U and D boundaries.    
For small $\alpha$, we see from Fig \ref{fig:irrev}(b) that the maximum of $\langle d_U \rangle$ is slightly lower than in the quasistatic limit. 
While the origin of this effect is not entirely clear, it is likely related to the fact that a small jump in concentration helps the system move beyond the quasistatic boundary and explore a slightly larger genotypic space. The trajectories would then be attracted preferentially towards a small number of nearby local maxima of high fitness, leading to these states evolving frequently along both the U and D boundaries. As $\alpha$ increases, a larger number of maxima become accessible through the adaptive walk, leading to a larger dispersion and reduced recurrence of the final state.         
}

\section{Summary and Discussion}

\rp{We have investigated a class of models of bacterial evolution under changing drug concentrations, and shown that their dynamics are closely related to that of driven disordered systems, exhibiting dynamical phenomena such as hysteresis and memory formation.  
As in the case of driven disordered systems, a state transition graph captures the dynamics of adaptive evolution in a changing environment. As a result, questions about evolution can be cast as questions about the graph topology. 
We have shown that in such models partial information about the changing environment is stored in evolving genotypes.
% and can be recovered from the state transition graph.
Although our analysis has focused on the relatively simple TIL model, we should emphasize that the memory effects described here are generic features of disordered systems.}

\st{Specifically, we} have found that the transitions between genotypes in a homogeneous population exhibit a number of generic properties which we have described in detail. \st{In particular, adaptive} walks are found to be short, \rp{{\it i.e.},} they involve a small number of mutations, when a slow change in the drug concentration renders a LFM unstable. \sgdnew{In this regime, the dynamics are not qualitatively altered by the precise nature of the adaptive walk.} Moreover, the systems generically exhibit hysteresis loops, {\em i.e.} a lack of reversibility when the direction of concentration change is reversed.
Conceptually, our work highlights the distinction between genotypic and phenotypic hysteresis.
Genotypic reversibility is found to be low for concentration changes with large jumps.
\sgd{However, the degree of phenotypic reversibility is found to be {{rather}} high under driving protocols involving both small and large jumps in the environmental parameter. Thus, the phenotypes are similar along both the upward and downward boundaries of drug concentration cycling. We have obtained asymptotic scaling approximations to phenotypes such as the number of mutations and the fitness and resistance levels as a function of drug concentration. These are relatively robust to modifications in the driving protocol, making them easily accessible to empirical testing.}

\sgd{Under slow concentration changes, we find that the genotypic hysteresis loops in several instances of our model exhibit perfect or near-perfect return-point memory (RPM), i.e. the capability to return to a previously visited genotype at which the direction of concentration change was reversed. Although not universally present, existence of RPM is relevant to applications such as the emergence of antibiotic resistance, since it implies that drug concentration cycling can lead to reversal of resistance. The degree of reversibility, including the presence of RPM, is affected by the distribution of the effects of mutations on the null-fitness and resistance phenotypes. However, empirical knowledge regarding the distribution of these mutational effects is still limited. Moreover, our model assumes as an approximation that the mutations combine non-epistatically, which rules out phenomena such as the occurrence of compensatory mutations that reduce the cost of resistance. Further work is needed to understand how various aspects of reversibility and memory are affected by these factors.}

\rp{
  From a broader perspective, our work introduces a systematic approach to understanding
  how the information about a changing environment
is encoded in the genotypes of a population. A possible direction for future research is the development of
algorithms that infer \mm{features of} the environmental history \sgd{from the knowledge of evolved genomes using the state transition graph}. 
Conversely,
our approach can be used to design treatment protocols that are optimized to avoid or slow down the evolution of drug
resistance by controlling the drug concentration or by cycling different antibiotics \cite{mira2015rational,nichol2015steering}.
% This problem will be pursued in future work.} 
}\\

\noindent {\bf Acknowledgments.} \jk{We wish to thank two anonymous reviewers for their useful comments on an earlier version
of the manuscript.}
SGD and JK acknowledge support by the Deutsche Forschungsgemeinschaft (DFG, German Research Foundation)
  within SFB 1310 \textit{Predictability in evolution}.
  MM was supported by the DFG under project No.~398962893, project No.~211504053 - SFB 1060, and
  under Germany’s Excellence Strategy - GZ 2047/1, project No.~390685813.}

%\showacknow{} % Display the acknowledgments section
%\bibliographystyle{unsrt}

\appendix

\section*{Appendix}
Here we include details of the mathematical models and computations used in our work. 

\section{Adaptive walks and changing environment}
\jk{Adaptive walks describe the evolutionary dynamics in a limiting regime of strong selection and weak mutation (SSWM)
  \cite{orr2002population,gillespie1984molecular}. Here \textit{strong selection} implies that only beneficial mutations can fix,
  and \textit{weak mutation} implies that no more than two genotypes are present in the  population at any time.}
% \sgd{Adaptive walks imply that we are in the regime of strong selection ({\em i.e.} only beneficial mutations can fix) and weak mutation ({\em i.e.} no more than two genotypes are present in the  population at any time).
  \sgd{The time-scale over which new beneficial mutations arise is $t_{\mbox{\small mut}}=1/(N\mu)$, where $N$ is the population size and $\mu$ is the mutation rate per genome per generation.  
The time-scale for fixation of a beneficial mutation is 
$t_{\mbox{\small fix}} = s^{-1} \ln(N^2 s)$ \cite{gerrish1998fate}, 
where $s>0$ is the selection coefficient of the beneficial mutation. The weak mutation condition requires that $t_{\mbox{\small fix}} \ll t_{\mbox{\small mut}}$, which implies
%requires that this time-scale must be much smaller than $t_{\mbox{\small mut}}$, implying
\begin{equation}
 N \mu \ln (N^2 s) \ll 1. \label{eq:awcond}
\end{equation}
In addition, we use the strong selection condition, where the fixation probability $1-e^{-2s}$ for beneficial mutations is derived from the well-known Kimura formula \cite{kimura1962probability} in the limit
\begin{equation}
 N s \gg 1.
\end{equation}
In our work, we have studied protocols in which we hold the concentration fixed for a waiting time $t_{\mbox{\small wait}}$ until a LFM is reached. Since this involves several evolutionary steps, a necessary condition on the waiting time is $t_{\mbox{\small wait}} \gg t_{\mbox{fix}}$.
  } \jk{This condition is approximately satisfied in experimental evolution protocols where the drug concentration is increased
    by a constant factor whenever the bacterial cell density has recovered a predetermined level, signaling that a new fitness
    maximum has been reached; see \cite{schenk2022population} for a recent example. The adaptive walk approximation requires in addition the condition \eqref{eq:awcond}, which was not the case in \cite{schenk2022population}, but could be achieved by working at smaller population
    sizes.}

\section{The TIL-Preisach equivalence}

\mm{We derive first some results for the TIL model which will be used in the following subsections. These results are general in the sense that they do not depend on} the shape of the dose-response function $w(x)$ as long 
as it satisfies the following properties:
\begin{itemize}

\item[(W1)] $w(x)$ is a continuous, strictly decreasing function for $x \ge 0$.

\item[(W2)] For all pairs of permissible values $(r,m)$ such that $r < 1$ and  $m> 1$, the curves $w(x)$ and $rw(x/m)$ intersect at precisely one point.

\end{itemize}  
As noted in \cite{das2020predictable} these conditions are satisfied for a large class of dose-response functions considered in the literature, including the exponential and half-Gaussian functions.  
When considering statistical results, we will specialize to the $n = 2$ Hill-type dose-response function $w(x) = 1/(1 + x^2)$. In this case,
in order for (W2) to be satisfied, the permissible pairs $(r,m)$ have to satisfy also $m^2r > 1$.
% , as is readily checked.

\mm{As given in the main text, the stability condition of a state of the TIL model is 
\begin{equation}
 \max_{j \in I^+[\vec{\sigma}] } \, \overline{x}_j  < \min_{i \in I^-[\vec{\sigma}] } \,  x_i,
 \label{eqn:genPartitionIneqSI}
\end{equation}
In the following, we shall assume that we are given a stable genotype $\vec{\sigma}$ such that
\begin{equation}
  x^-[\vec{\sigma}] = m_{\vec{\sigma}} \, \overline{x}_\ell \quad \mbox{and} \quad 
  x^+[\vec{\sigma}] = m_{\vec{\sigma}} \, x_u, \label{eqn:xminusplus_ellk} 
\end{equation}
hold with $x^-[\vec{\sigma}] < x^+[\vec{\sigma}]$, implying that the stability condition Eq. \ref{eqn:genPartitionIneqSI} is satisfied, and the sites $\ell$ and $u$ are those at which the maximum, respectively minimum of the terms on the left and right hand side of the inequality are attained. We say that sites $\ell$ and $u$ are the least stable sites in  the sense that the first mutation will occur there under concentration decreases or increases, respectively. The stability range of $\vec{\sigma}$ is then given as 
$(x^-[\vec{\sigma}],x^+[\vec{\sigma}])$.} 

\mm{By setting $h^-_i = \overline{x}_i$ and $h^+_i = x_i$, the condition \eqref{eqn:genPartitionIneqSI} becomes the stability condition for hysteron configuration $\vec{\sigma}$ of the Preisach model, as derived in the main text. The stability range  
$(h^-[\vec{\sigma}], h^+[\vec{\sigma}])$ is then given by 
\begin{equation}
  h^-[\vec{\sigma}] =  \overline{x}_\ell \quad \mbox{and} \quad 
  h^+[\vec{\sigma}] =  x_u, \label{eqn:xminusplus_ellkPR} 
\end{equation}
which formally can be obtained from the stability range of its corresponding TIL state, \eqref{eqn:xminusplus_ellk}, by setting $m_{\vec{\sigma}} = 1$. Thus while a TIL model and its Preisach analogue have identical sets of stable states, the stability ranges of 
the two are different in general.}

\subsection{The proximity-property of stable states}

\mm{Given a stable TIL-Preisach state $\vec{\sigma}$, we let again $\ell$ and $u$ denote its least stable sites. The following result 
holds for the stability ranges $h^\pm[\vec{\sigma}]$ of the Preisach model:
\begin{align}
  h^-[\vec{\sigma}] \le h^-[\vec{\sigma}^{+u}]  &< h^+[\vec{\sigma}] < h^+[\vec{\sigma}^{+u}], \label{eqn:AQSPr1} \\
 h^-[\vec{\sigma}^{-\ell}] < h^-[\vec{\sigma}] &< h^+[\vec{\sigma}^{-\ell}] \le h^+[\vec{\sigma}]. \label{eqn:AQSPr2}
\end{align}
Note that these inequalities establish in particular that  $h^-[\vec{\sigma}^{+u}] < h^+[\vec{\sigma}^{+u}]$ and 
$h^-[\vec{\sigma}^{-\ell}] <  h^+[\vec{\sigma}^{-\ell}]$, so that $\vec{\sigma}^{+u}$ and $\vec{\sigma}^{-\ell}$ are 
stable Preisach states as well. But by the TIL-Preisach equivalence they must be stable TIL states, too. This can also be seen 
by noting that the corresponding stability ranges are obtained from those of its Preisach analogue by multiplication by $m_{\vec{\sigma}^{+u}}$, respectively $m_{\vec{\sigma}^{-\ell}}$. This is the proximity-property of the stable states for the TIL model and its Preisach 
equivalent.}

\mm{We only prove the inequalities in \eqref{eqn:AQSPr1}, the proof of \eqref{eqn:AQSPr2} is similar. First note that $I^+[\vec{\sigma}]$ is a proper subset of $I^+[\vec{\sigma}^{+u}]$, and likewise $I^-[\vec{\sigma}^{+u}]$ is a proper subset of $I^-[\vec{\sigma}]$. Since each of the two sets  $\{h^\pm_i\}_{i=1}^N$ is assumed to have distinct elements, it follows that 
\begin{equation}
 \min_{ j \in I^-[\vec{\sigma}^{+u}]} \, h^+_j  > \min_{ j \in I^-[\vec{\sigma}]} \, h^+_j.
\end{equation}
%because the minimum on the left-hand side is taken over the proper subset of $I^-[\vec{\sigma}]$, which is obtained  by removing from it the element $j = u$ with lowest $h^+_j$. 
Thus $h^+[\vec{\sigma}^{+u}] > h^+[\vec{\sigma}]$ and the rightmost inequality of \eqref{eqn:AQSPr1} has been proven. Next, consider $h^-[\vec{\sigma}^{+u}]$ and particularly
its least stable element $k$ under field decreases. Recall that we denoted the corresponding element for  $\vec{\sigma}$ as $\ell$. 
Therefore, either (i) $k = \ell$ or (ii) $k = u$. 
In the former case, it must have been that $h^-_u < h^-_\ell$ and therefore $h^-[\vec{\sigma}^{+u}] =h^-[\vec{\sigma}] = h^-_\ell$. In the letter case, the opposite must be true, i.e. $h^-_u > h^-_\ell$, and therefore $h^-[\vec{\sigma}^{+u}] = h^-_u > h^-[\vec{\sigma}]$. Combining these two cases, it follows that 
$h^-[\vec{\sigma}^{+u}] \ge h^-[\vec{\sigma}]$, thereby establishing the left most inequality of \eqref{eqn:AQSPr1}. However, since by definition $h^-_u < h^+_u$, in both cases it must be that $h^-[\vec{\sigma}^{+u}]  < h^+[\vec{\sigma}] = h^+_u$, thereby establishing the middle inequality of \eqref{eqn:AQSPr1}.}

\subsection{Construction of stable states from the symbolic order sequence $p$}

\mm{Here we provide an explicit construction of the set of stable states of the TIL model (and its Preisach equivalent) 
from the symbolic order sequence $p$, which represents the ordering of the $2L$ concentrations $\overline{x}_i$ and $x_j$. 
Denoting the set of stable states associated with $p$ as $\mathcal{S}_p$, this set can be partitioned into the 
subset of states $\mathcal{S}_{p,u}$ whose least stable site under concentration increases is $u$, with $u = 1, 2, \ldots, L$ (this subdivision leaves out the all-mutant $\vec{\sigma} = \vec{1}$, which is always stable, and we assign it to the singleton set $\mathcal{S}_{p,L+1}$).}  

\mm{To illustrate the construction of $\mathcal{S}_{p,u}$, consider the 
%ordering of the $\overline{x}_i$ and $x_j$ along with the corresponding 
order sequence \eqref{eqn:pexample} 
of the example given in the main text: 
\begin{align}
\overline{x_1} < \overline{x_2} < x_1 < \overline{x_5} < \overline{x_3} &< x_2 < \overline{x_4} <  x_3  < x_4  < x_5, \nonumber \\
&\Leftrightarrow \\
%\label{eqn:porder}
p = \quad \overline{1} \quad  \overline{2} \quad 1 \quad  &\overline{5} \quad \overline{3} \quad 2 \quad \overline{4} \quad 3 \quad 4 \quad 5.
\nonumber
\end{align}
}

\mm{Let us construct $\mathcal{S}_{p,2}$, the set of stable states with least stable site $u = 2$. All stable states must satisfy the inequality \eqref{eqn:genPartitionIneqSI}, and in particular the right hand side of it must be equal to $x_2$. 
This can only be the case, if $\sigma_2 = 0$ and $\sigma_1 = 1$, i.e. the site $1$ must belong to $I^+[\vec{\sigma}]$. In terms of the order sequence $p$ this condition is equivalent to requiring that any element $i$ without an overbar which is located to the left of element $u$ 
must be assigned as $\sigma_i = 1$. }

\mm{Likewise, in order to ensure that the left hand side of the inequality \eqref{eqn:genPartitionIneqSI} is strictly less than 
its right hand side, we require that any site $j$ with an overbar to the right of element $u$ must have $\sigma_j = 0$. In our example 
this requires that $\sigma_4 = 0$. Any site $k$ left undetermined by these two conditions can be assigned as $ \sigma_k = 0$ or $1$. In 
the above example these conditions leave the sites $k = 3$ and $5$ undetermined so that $\mathcal{S}_{p,2}$ has four elements given by 
$\mathcal{S}_{p,2} = \{ (10000), (10100), (10001), (10101) \}$. Repeating the construction for all values of $u$, 
the reader may verify that $\mathcal{S}_p$ has $14$ states in total. These are the states shown in Fig.~\ref{fig:trgraphs}.}  

\mm{In the following we will make repeated use of two results whose validity is a direct consequence of the construction of stable states given above: (i) Given a symbolic ordering sequence $p$ and using Eq. \ref{eqn:genPartitionIneqSI}, the set of all stable genotypes $\vec{\sigma}$ can be inferred from it, and hence the possible pairs of least stable sites $(\ell,u)$ associated with these. (ii)
Given any pair of sites $\ell$ and $u$ such that $\overline{x}_\ell < x_u$, there exists a symbolic ordering sequence $p$ such that Eq. \ref{eqn:xminusplus_ellk} holds and thus
$\ell$ and $u$ are the least stable sites for some stable state $\vec{\sigma}$.} 
Whenever we assume that Eq. \ref{eqn:xminusplus_ellk} holds, this  will either imply that we are given a specific order sequence $p$ and that with respect to $p$ the state $\vec{\sigma}$ is stable with $(\ell,u)$ being the pair of least-stable sites, or alternatively, we are given $(\ell,u)$ and consider the set of order sequences $p$ and stable states $\vec{\sigma}$ compatible with this choice of least stable sites. The particular point of view will be clear from the context.    

\subsection{Properties of secondary mutations}
\subsubsection{Secondary mutation must be complementary} 
We first derive some simple inequalities {for the TIL} model. \mm{The two properties given below follow immediately from}  assumptions (W1) and (W2) made above for $w(x)$:

\begin{align}
 x &< x_i \quad \Leftrightarrow \quad w(x) > r_i w \left ( \frac{x}{m_i} \right ), \label{eqn:ineqxless} \\
 x &> x_i \quad \Leftrightarrow \quad w(x) < r_i w \left ( \frac{x}{m_i} \right ).\label{eqn:ineqxgreater}
\end{align}

Now, let ${\vec{\sigma}}$ be a stable state, such that Eq. \ref{eqn:xminusplus_ellk} holds. 
 Then, we can show from the previous results that for all $i \in I^-[\vec{\sigma}] \setminus \{u\}$, 
     \begin{equation}
        \frac{ f_{\vec{\sigma}^{+u,+i}} (x^+[\vec{\sigma}]) } { f_{\vec{\sigma}^{+u}}(x^+[\vec{\sigma}]) } 
 < 1,
  \end{equation}
   and for all $j \in I^+[\vec{\sigma}] \setminus \{\ell \}$,
     \begin{equation}
        \frac{ f_{\vec{\sigma}^{-\ell,-j}} (x^-[\vec{\sigma}])} { f_{\vec{\sigma}^{-\ell}}(x^-[\vec{\sigma}])}  < 1.
   \end{equation}
The last two inequalities together assert that first secondary mutations which are in the same direction as the primary mutation, are fitness-decreasing. Thus either the first mutation leads to 
a LFM, and hence there will be no further mutations, or the first secondary mutation must be complementary to the original mutation.

\subsubsection{Locations of complementary secondary mutations} 
Let ${\vec{\sigma}}$ be a stable state such that Eq. \ref{eqn:xminusplus_ellk} holds. Then for $i \in I^+[\vec{\sigma}]$, 
     \begin{align}
\overline{x}_i &> \overline{x}_u  \quad \Leftrightarrow \quad     \frac{ f_{\vec{\sigma}^{+u,-i}} (x^+[\vec{\sigma}]) } { f_{\vec{\sigma}^{+u}}(x^+[\vec{\sigma}]) } 
 > 1. \label{eqn:fpkmi}
  \end{align}
Therefore, subsequent to an initial mutation under concentration increase at site $u$,   
fitness increasing complementary mutation sites are those sites $i \in I^+[\vec{\sigma}]$ for which the symbol $\overline{i}$ is 
% sites $i$ of fitness increasing complementary secondary mutations correspond to the sites $\overline{i}$ 
located to the right of $\overline{u}$ in the order sequence $p$. Note in particular, that the initial mutation site $u$ itself cannot be also the site for a subsequent secondary mutation, as this would have implied that $\vec{\sigma}$ has a higher fitness than $\vec{\sigma}^{+u}$ at the triggering concentration. 

Likewise, for  $j \in I^-[\vec{\sigma}]$,
     \begin{align}
x_j &< x_\ell \quad \Leftrightarrow \quad  \frac{ f_{\vec{\sigma}^{-\ell,+j}} (x^-[\vec{\sigma}])} { f_{\vec{\sigma}^{-\ell}}(x^-[\vec{\sigma}])}  > 1. \label{eqn:fmellpj}
   \end{align}
Any secondary mutation following an initial mutation under concentration decrease at site $\ell$, must be a site $j \in I^-[\vec{\sigma}]$ located to the left of $\ell$ in the symbolic order sequence $p$. The statements Eq. \ref{eqn:fpkmi} and Eq. \ref{eqn:fmellpj} are proven by 
repeated application of  Eq. \ref{eqn:ineqxless}, Eq. \ref{eqn:ineqxgreater}, and the properties of ordering sequences $p$ that are compatible with the assumption Eq. \ref{eqn:xminusplus_ellk}. 

\subsubsection{Secondary mutations cannot cause transitions to a subset or superset}
Assume the contrary. Then, according to the property of Directed Path Accessibility, a path must exist where the first secondary mutation is in the same direction as the original mutation. But this is not possible according to the previous result. 

\subsubsection{Conditions for the absence of secondary mutations}
Consider a realization of the TIL model with $L$ sites and let $\vec{\sigma}$ be a stable state satisfying Eq. \ref{eqn:xminusplus_ellk}. 
Assume that we are given a symbolic ordering sequence $p$ compatible with Eq. \ref{eqn:xminusplus_ellk}. For 
any site $u = 1, 2, \ldots L$, we will be interested in the interval of elements of $p$ that is bounded to the left by $\overline{u}$ and to the right by $u$.
Denote by $\mathcal{I}_u$ the set of sites $j$  that appear in this interval without overbars. 
Likewise, let $\overline{\mathcal{I}}_u$ be the set of sites that appear in this interval with overbars. Our definition is such that neither of the two sets of sites $\mathcal{I}_u$ and $\overline{\mathcal{I}}_u$  contain $u$. 

Consider now transitions out of $\vec{\sigma}$ under concentration increases.
By assumption, under a concentration increase to (a value slightly above) $x^+[\vec{\sigma}]$, the site $u$ will mutate first, $\sigma_u = 0 \to 1$, 
leading to $\vec{\sigma}^{+u}$,
and as a result, the upper limit of the 
stability range of $\vec{\sigma}^{+u}$ increases to 
$x^+[\vec{\sigma}^{+u}] > x^+[\vec{\sigma}]$. In order to assert the stability of $\vec{\sigma}^{+u}$ at the concentration $x^+[\vec{\sigma}]$
which triggered the mutation at $u$, 
we must require that 
\begin{equation}
x^-[\vec{\sigma}^{+u}] \le  x^+[\vec{\sigma}].
\label{eqn:comp}
\end{equation}
If this condition is not satisfied, then $x^\pm[\vec{\sigma}^{+u}] > x^+[\vec{\sigma}]$ and at least one secondary mutation occurs. We thus need to find conditions under which Eq. \ref{eqn:comp} holds. 
 
Now in terms of the ordering sequence $p$ , the site $\overline{\ell}$ must be located to the left of $u$, as must be the site $\overline{u}$.
Moreover, since $\ell$ and $u$ have to be distinct, 
$\overline{\ell}$ is either to the left or right 
of $\overline{u}$. In the former case we have $\overline{x}_\ell < \overline{x}_u$, and hence 
\begin{equation}
  x^-[\vec{\sigma}^{+u}] = m_{\vec{\sigma}^{+u}} \overline{x}_u = m_{\vec{\sigma}} x_u = x^+[\vec{\sigma}].
\end{equation}
Since Eq. \ref{eqn:comp} is satisfied, genotype $\vec{\sigma}^{+u}$ is a local fitness maximum at this concentration and 
there will therefore  be no secondary mutations. 
Suppose next that $\overline{x}_\ell > \overline{x}_u$. In this case 
\begin{equation}
x^-[\vec{\sigma}^{+u}] = m_{\vec{\sigma}^{+u}} \, \overline{x}_\ell = m_{\vec{\sigma}}  \,\frac{\overline{x}_\ell}{\overline{x}_u} \, x_u > x^+[\vec{\sigma}].
\end{equation}
Therefore there will be at least one complementary secondary mutation at some site $i \in I^+[\vec{\sigma}]$. 
Condition Eq. \ref{eqn:fpkmi} asserts that in order for such a mutation to be fitness increasing, 
$i$ must be such that $\overline{x}_i > \overline{x}_u$. Using Eq. \ref{eqn:genPartitionIneqSI}, it \mm{follows} 
that the stability condition of $\vec{\sigma}$, as given by Eq. \ref{eqn:xminusplus_ellk}, implies that $\overline{x}_i < x_u$, so 
that the secondary mutation site must be contained in the set $\overline{\mathcal{I}}_u$. Note in particular that 
the site $\ell$ itself satisfies these conditions and hence is a possible candidate for the first secondary mutation. 

Combining all of the above results, under increasing concentration, a secondary mutation will occur, if and only if 
the set $\overline{\mathcal{I}}_u$ is non-empty, and the state $\vec{\sigma}$ is such that, for some
 $j \in \overline{\mathcal{I}}_u$ we have $\sigma_j = 1$. Conversely, a secondary mutation will {\em not} occur if and only if
one of the following two conditions holds:
 \begin{itemize}
    \item [(U1)] The set $\overline{\mathcal{I}}_u$ is empty. 
    \item [(U2)] The set $\overline{\mathcal{I}}_u$ is non-empty, and the state $\vec{\sigma}$ is such that, for each $j \in \overline{\mathcal{I}}_k$ we have $\sigma_j = 0$.  
 \end{itemize}

 In a similar manner, one can show that under decreasing concentration a secondary mutation will not occur, if and only if one of the following two
 conditions holds:
 \begin{itemize}
    \item [(D1)] The set $\mathcal{I}_\ell$ is empty. 
    \item [(D2)] The set  $\mathcal{I}_\ell$ is non-empty, and the state $\vec{\sigma}$ is such that, for each $j \in \mathcal{I}_\ell$ we have $\sigma_j = 1$.  
 \end{itemize}

Observe now that in order for secondary mutations to be absent from all transitions in a TIL model, the sets $\mathcal{I}_k$ and $\overline{\mathcal{I}}_k$ have to be empty for each $k = 1, 2, \ldots, L$, since otherwise there will exist stable states for which conditions 
(U2) or (D2) can be made not to hold. \mm{The} only ordering sequence for which both of these 
sets are empty \mm{for each $k$} is the sequence 
    \begin{equation}
     p = \overline{1} \quad 1 \quad \overline{2} \quad 2 \quad \dots \quad \overline{L} \quad L. \nonumber
    % \label{eqn:pnosecndmut}
    \end{equation}

\section{Statistical Results}
\mm{In the following subsections, were derive the statistical results discussed in the main body of the paper.}

\subsection{Probability density function used in the numerics}
We assume that the dose-response curve is of Hill-type with $n = 2$. In order to satisfy the requirement (W2) for the dose-response function, 
the parameters $(r_j,m_j)$ must be chosen such that $m_j^2 r_j > 1$ for each $j = 1, 2, \ldots, L$. We further assume that the pairs $(r_j,m_j)$ are independently and identically distributed, so that their 
joint density is given by $Q(\{ r_i, m_i \}) = \prod_{j=1}^L P(r_j,m_j)$. 
We write $P(r_j,m_j)=P_1(r_j) P_2(m_j|r_j)$. We chose 
\begin{eqnarray}
 P_1(r) &=& \sqrt{\frac{2}{\pi}}\frac{e^{-\frac{(\ln r)^2}{2}}}{r} \label{eq:dist1} \\
 P_2(m|r) &=& \mathcal{N}~ \frac{e^{-\frac{(\ln m)^2}{2}}}{m} \Theta\big(m-\frac{1}{\sqrt{r}}\big) \label{eq:dist2},
\end{eqnarray}
where $\Theta(\cdot)$ is the Heaviside step function, and $\mathcal{N}$ is the appropriate normalization constant. This choice is for easy of implementation. A similar model was used in \cite{das2020predictable}. 
%We have sampled each of the parameter pairs $(r_j,m_j)$ using the following 
%rejection method. Letting $y=-\ln r$ and $z=\ln m$, we first draw $y$ from a half-gaussian distribution with %density $P_y(y) = \sqrt{{2}{\pi}}e^{-y^2/2}$, $y\ge 0$. We then keep drawing $z$ values  independently from the %same half-gaussian distribution until condition (W2) is satisfied, {\em i.e.} $2z - y > 0$ 
%  We have chosen 
%  $P(r,m) \sim P_r(r)P_m(m)\Theta(m-1/\sqrt{r})$, where $\Theta$ is the Heaviside function. The  density $P_r$  is chosen such that $y=-\ln r$ is distributed as a half-Gaussian, i.e. $P_y(y) = \sqrt{{2}{\pi}}e^{-y^2/2}$, $y\ge 0$. 
%  Similarly, the density $P_m$ is chosen such that $z=\ln m$ satisfies the same distribution, i.e. $P_z(z) = \sqrt{{2}{\pi}}e^{-z^2/2}$, $z\ge 0$.

\subsection{Asymptotic number of stable states}
Consider a genotype $\vec{\sigma}$ with $n$ mutations, i.e. $\sum_i \sigma_i = n$. The number of such genotypes is 
 $\binom{L}{n}$ {and} such a $\vec{\sigma}$ is a stable state if Eq. \ref{eqn:genPartitionIneqSI} holds. Since $x_i$ and $\bar{x}_i$ are independent for distinct sites, the probability density that the left hand side of Eq. \ref{eqn:genPartitionIneqSI} is less than $z$ and the right hand side is greater than $z$ is 
 {$\frac{d}{dz} F^{L-n}_{\bar{x}}(z) [1-F_x(z)]^n$}. Then the mean number of stable states is 
 \begin{eqnarray}
  \langle N_{ss} \rangle &=& \sum_{n=0}^L  
  {\left (
   \hspace{-0.3em} 
  \begin{array}{l} L \\ n \end{array} 
  \hspace{-0.3em} 
  \right )} 
  \int dz~\left [\frac{d}{dz} F^{L-n}_{\bar{x}}(z)  \right ] \left [1-F_x (z) \right ]^n  \nonumber \\ 
  &=& L \int dz \, {\left [1-F_x(z)+F_{\bar{x}}(z) \right ]^{L-1} \, {F'_{\bar{x}}(z)}},
 \end{eqnarray}
from which the result in the main text follows using a saddle point approximation for $L$ large.

\subsection{Asymptotic approximation for number of mutations}
The fitness $f$ of a genotype ${\vec{\sigma}}$ can be expressed as
{
\begin{equation}
 \ln f = \sum_i \sigma_i \ln r_i -\ln \left ( 1+ x^2 \, e^{-2 \, \sum_i \sigma_i \ln m_i} \right)
 \label{eq:logfit}
\end{equation}
}
The number of mutations in the genotype is $n=\sum_i \sigma_i$. A simple heuristic that produces good approximations for the mean of various quantities at large $L$ is a s follows: we consider the fitness of a genotype to be a function of $x$ and $n$ only, and replace the parameters associated with the mutations by suitable averages. Thus, we write Eq. \ref{eq:logfit} as 
{
\begin{equation}
 \ln f(n) \simeq  -n a - \ln \left ( 1+ x^2 \,  e^{-2nb } \right ),
 \label{eq:fapprox}
\end{equation}
}
where $a=-\langle\ln r \rangle$ and $b= \langle \ln m \rangle$.
For any given $x$, one can now maximize Eq. \ref{eq:fapprox} with respect to $n$, yielding an approximation to the mean mutation number at $x$ for stable maxima. Taking the derivative of the above with respect to $n$ and setting it to zero produces the equation: 
{
\begin{equation}
\frac{2b x^2 e^{-2nb}}{1+x^2 e^{-2nb}}=a.\nonumber
\end{equation}
 The solution to this is 
\begin{equation}
 n = \frac{\ln x}{b}+\frac{1}{2b} \ln \left ( \frac{2b}{a}-1 \right ). 
\end{equation}
}
For large $x$ and therefore large $n$, the leading order is 
\begin{equation}
 n \simeq \frac{\ln x}{b}. 
 \label{eq:nx}
\end{equation}
This estimate works well when $L$ is large and $ 1 \ll \frac{\ln (x)}{\langle \ln m \rangle} \ll L $.

\bibliography{TPbib}

%\matmethods{
%Example text for subsection.
%}

%\showmatmethods{} % Display the Materials and Methods section

\end{document}